\documentclass[aps,pre,twocolumn,showpacs,superscriptaddress,groupedaddress]{revtex4}
\usepackage{graphicx}
\usepackage{amsfonts}
\usepackage{bm}%
\usepackage[colorlinks=true,linkcolor=blue,filecolor=blue,menucolor=yellow,urlcolor=blue,
citecolor=blue,anchorcolor=blue]{hyperref}
\usepackage{graphicx}% Include figure files
\usepackage{dcolumn}% Align table columns on decimal point
\usepackage{rotating}
\usepackage{subfig}
\usepackage[normalem]{ulem}
\usepackage{color}

\usepackage{amsmath}
\usepackage{ulem}
\usepackage{filecontents}
\usepackage[titletoc,toc,title]{appendix}
\usepackage[font=small,labelfont=bf,
   justification=justified,
   format=plain]{caption}

\usepackage[export]{adjustbox}
\usepackage{floatrow}

\begin{document}

\title{Far from equilibrium dynamics of tracer particles embedded in a growing multicellular spheroid}
\author{Himadri S. Samanta}\affiliation{Department of Chemistry, University of Texas at Austin, TX 78712}
\author{Sumit Sinha}\affiliation{Department of Physics, University of Texas at Austin, TX 78712}
\author{D. Thirumalai}\affiliation{Department of Chemistry, University of Texas at Austin, TX 78712}

%\affiliation{Institute For Physical Science and Technology, University of Maryland, College Park, MD 20742}

%\begin{document}
%\maketitle

\begin{abstract}
Local stresses on the cancer cells (CCs) have been measured by embedding inert tracer particles (TPs) in a growing multicellular spheroid. The utility of the experiments requires that the TPs do not alter the CC microenvironment.  We show, using theory and extensive simulations, that proliferation and apoptosis of the CCs, drive the dynamics of the TPs far from equilibrium. 
On times less than the CC division times, the TPs exhibit sub-diffusive behavior (the mean square displacement, $\Delta_{TP}(t) \sim t^{\beta_{TP}}$ with $\beta_{TP}<1$).  Surprisingly, in the long-time limit, the motion of the TPs is {\it hyper-diffusive} ($\Delta_{TP}(t) \sim t^{\alpha_{TP}}$ with
$\alpha_{TP}>2$) due to persistent directed motion for a number of CC division times.  In contrast, CC proliferation  randomizes their motion resulting from jamming at short times  to super-diffusive behavior, with $\alpha_{CC}$ exceeding unity, at long times.  Surprisingly, the effect of the TPs on CC dynamics and radial pressure is negligible, suggesting that the TPs are reliable reporters of the CC microenvironment.   %Our predictions are testable using imaging experiments.
\end{abstract}

\date{\today}
\maketitle
%\doublespacing

%\section{Introduction}
%Self-organized active soft matter {exhibiting} collective motion, instabilities, patterns, and dynamic disorder, is one of the most spectacular manifestations of living systems. Well known examples of such systems include schools of fish, flocks of birds, aggregations of molecular motors, remodeling of growing tissues, etc.\cite{Marchetti13RMP,Ram10ARCMP, Vicsek95PRL,Kruse01PRL,Gregoire04PRL,Johann12PRL,Toner95PRL,cates10PNAS,Basan11PRL,Montel11PRL,Toner12PRL,Chen13PRL}. 

The interplay between  short-range forces and non-equilibrium processes arising from cell division and apoptosis  gives rise to unexpected dynamics in the collective migration of cancer cells \cite{Friedl09NatRevMolCellBiol,Shaebani20NatRevPhys,kumar2009mechanics,desoize2000multicellular, walenta2000metabolic, laurent2013multicellular, valencia2015collective}. An example is the invasion of cancer cells (CCs) in a growing multicellular spheroid (MCS), which is relevant in cancer metastasis \cite{laurent2013multicellular, valencia2015collective}. Imaging experiments show that collective migration of a group of cells that maintain contact for a long period of time exhibits far from equilibrium characteristics \cite{valencia2015collective, richards20184d,martino2019wavelength, han2019cell, palamidessi2019unjamming}. 
Simulations and theory explain some of the experimental observations \cite{Abdul17Nature,Himadri18PRE,sinha2019spatially}. Dynamics in a growing MCS is  reminiscent of the influence of active forces in abiotic systems \cite{marchetti2013hydrodynamics, bechinger2016active,Nandi18PNAS}. In a growing MCS the analogue of active forces are self-generated \cite{sinha2020self}, arising from biological events characterized here by cell division and apoptosis.

Experiments that probe the local stresses or pressure on the CCs \cite{rauzi2011cortical,hutson2003forces,boucher1990interstitial,fadnes1977interstitial,campas2014quantifying,Dolega17NC} have provided insights into the mechanism by which the CCs invade the extracellular matrix. Recently, the stresses within MCSs were measured \cite{Dolega17NC} by embedding micron-sized inert deformable polyacrylamide beads as probes. {\color{black} For a TP to be a reliable sensor, it should not alter the dynamics of the CCs and the microenvironment, as assessed by local pressure. However, the influence of TPs on the CCs is unknown -- a gap that we fill here.} 

%Inspired in part by this experiment, we used theory and simulations to {\color{blue}probe the dynamics of TPs  in the MCS and the impact on the tumor}. %In our model, a CC grows till it reaches a critical size whereupon it divides into two at a rate $k_b$. In addition, a randomly chosen CC could undergo apoptosis at a rate, $k_a$. The inequality ($k_a << k_b$), chosen to mimic tumor growth, gives rise to self-generated forces, driving the system far from equilibrium \cite{doostmohammadi2015celebrating}.
%We model the CCs and the TPs as  deformable objects subject to short-ranged repulsive elastic forces and adhesive interactions arising from other CCs and TPs. We investigated the relevant continuum description of the collective behavior of a colony of tracers in the finite as well as in the long time limit using theory~\cite{Parisi81ES,Himadri06PLA,Himadri06PRE} and simulations~\cite{Abdul17Nature}. 

We used theory and simulations to {\color{black}probe the dynamics of the TPs  in the MCS, and the impact of the TPs on the CC mobility as well as radial pressure}. The central results are :~(1)~The TPs exhibit sub-diffusive motion in the intermediate time scale ($t\lesssim\frac{1}{k_b}$), with the mean squared displacement, $\Delta_{TP}(t) \sim t^{\beta_{TP}}$, with $\beta_{TP}$ less than unity. In the long time limit, ($t\gtrsim\frac{1}{k_b}$), the TPs undergo {\it hyper-diffusive} motion, $\Delta_{TP}(t) \sim t^{\alpha_{TP}}$ with $\alpha_{TP}$ greater than 2. In contrast, the CCs, which are jammed at $t\lesssim\frac{1}{k_b}$, exhibit {\it super-diffusive} dynamics.
%(2) In the long time limit, the extent of migration of the TPs relative to CCs is greater ($\alpha_{TP}^T=2.28 > $\alpha_{CC}^T$.   The values of both $\alpha_{TP}^S (=2.3)$ and $\alpha_{CC}^S (=1.47)$ (superscript $S$ stands for simulation), obtained using simulations  are in quantitative agreement with the theoretical predictions. 
(2)  The dynamics of the CCs in the intermediate time regime changes slightly as the size of the TPs are varied. {\color{black} Remarkably, the long time exponent ($\alpha_{CC}$) as well as the decrease in the pressure as a function of the distance from the center of the MCS remains unaffected in the presence of TPs.} %{\color{blue} Thus, TPs serve as good reporters of physical measurements (eg. stress) inside tumors since the dynamical properties of the medium remains unaltered.} % which in principle is possible using imaging experiments.

\begin{figure}
{\includegraphics[width=0.8\linewidth] {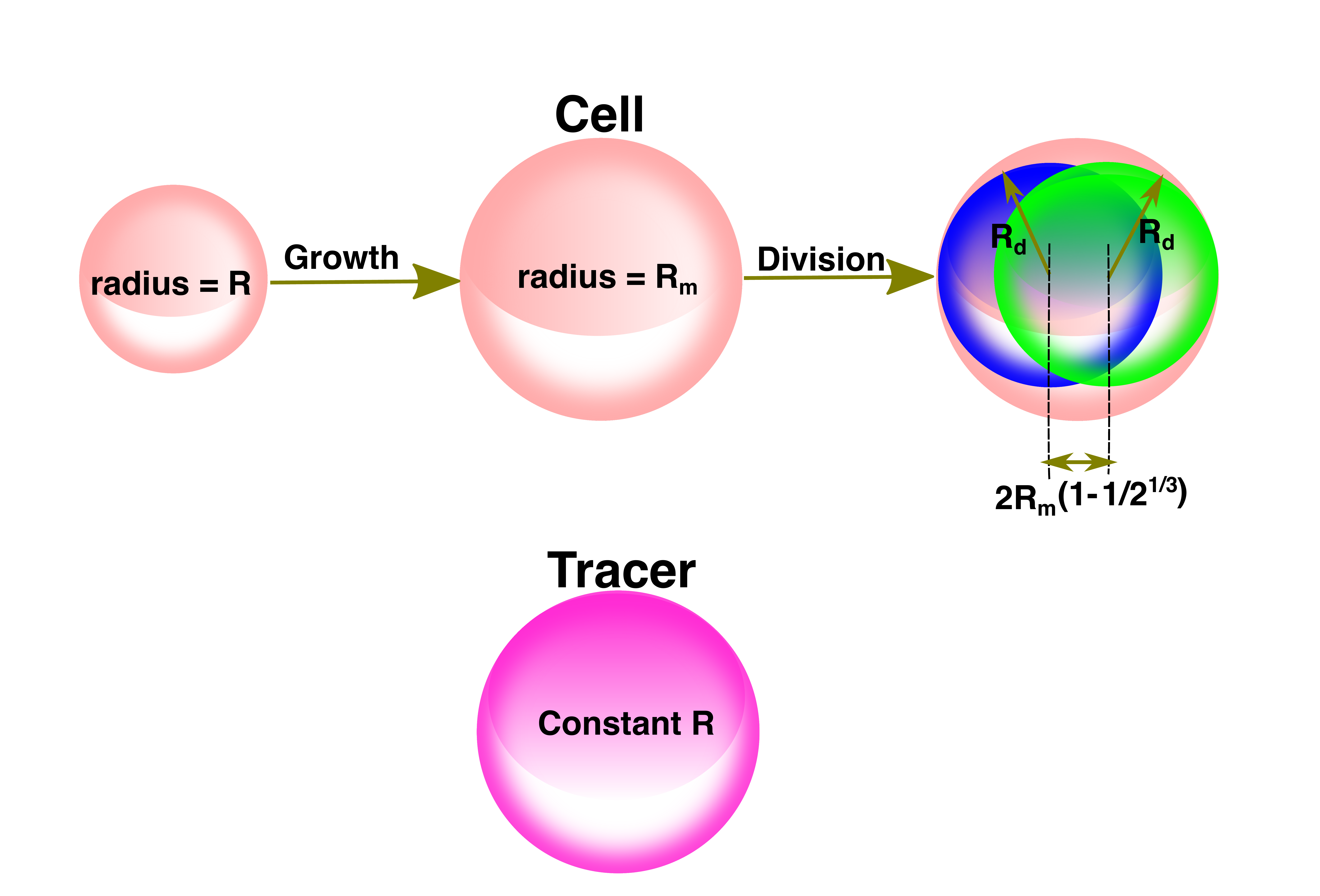}} 
\vspace{-.1 in}
\caption{Difference between the CCs (pink sphere) and the TPs (magenta sphere). Cell division creates two daughter CCs (blue and green sphere), each with $R_d=\frac{R_m}{2^{1/3}}$. Their relative position is displaced by $2R_m(1-\frac{1}{2^{1/3}})$ with the orientation being random with respect to the sphere center. }
\label{tracer_cell}
\end{figure}

{\it{Theory}}:{\color{black}
~We consider the dynamics of the TPs in a growing MCS in a dissipative environment by neglecting inertial effects.
 %We assume that the TP and CC dynamics are governed by stochastic equations. 
 The deformable but inert TPs, do not grow, divide or undergo apoptosis. 
 %experience systematic short-range volume exclusion and attractive interactions arising from other TPs and CCs. 
 The CCs grow and divide at a given rate, and undergo apoptosis (see Figure \ref{tracer_cell}). The CCs and TPs experience systematic short-ranged attractive and repulsive forces arising from other CCs and TPs \cite{Abdul17Nature}. 
To obtain the dynamics of the TPs and CCs, we introduce the density fields $\phi({\bf r},t)=\sum_i \delta[{\bf r}-{\bf r}_i(t)]$ for the CCs, and $\psi({\bf r},t)=\sum_i \delta[{\bf r}-{\bf r}_i(t)]$ for the TPs.  A formally exact Langevin equation for $\phi({\bf r},t)$}, and $\psi({\bf r},t)$ may be derived using the Dean's method~\cite{Dean96JPA} accounting for diffusion and non-linear interactions.

 \floatsetup[figure]{style=plain,subcapbesideposition=top}

To model a growing MCS, we modify the density equation for the CCs phenomenologically by adding a non-linear source term, $\propto \phi(\phi_0-\phi)$, accounting for cell birth and apoptosis, and a non-equilibrium noise term that breaks the CC number conservation. 
The noise, $f_\phi$, satisfies $<f_\phi({\bf r},t)f_\phi({\bf r'},t')>=\delta({\bf r}-{\bf r}')\delta(t-t')$.
The source term, $\propto \phi(\phi_0-\phi)$, represents the birth and apoptosis, with  $\phi_0=\frac{2k_b}{k_a}$\cite{Doering03PA,Gelimson15PRL}.
The coefficient ,~$\sqrt{k_b \phi+\frac{k_a}{2}\phi^2} $ of $f_\phi$, is the noise strength, corresponding to number fluctuations.

The $\phi({\bf r},t)$ and $\psi({\bf r},t)$ fields obey,
\begin{eqnarray}
\frac{\partial \psi({\bf r},t)}{\partial t}&=&D_\psi \nabla^2 \psi({\bf r},t)+ 
{\bf \nabla }\cdot \left(\psi({\bf r},t){\bf J}\right)+\tilde{\eta}_\psi , \label{trdensity}\\
\frac{\partial \phi({\bf r},t)}{\partial t}&=&D_\phi \nabla^2 \phi({\bf r},t)+ 
{\bf \nabla }\cdot \left(\phi({\bf r},t){\bf J}\right) +\nonumber\\
&&
\frac{k_a}{2} \phi(\frac{2k_b}{k_a}-\phi)+\sqrt{k_b \phi+\frac{k_a}{2} \phi^2} f_\phi+\tilde{\eta}_\phi
 \,
\label{phi10}
\end{eqnarray}
%\end{widetext}
% \vspace{-.2 in}
where ${\bf J}=\int_{\bf r'}[\psi({\bf r'},t)+ \phi({\bf r'},t)]{\bf \nabla}U({\bf r-\bf{r'}})$, $\tilde{\eta}_\psi({\bf r},t)={\bf \nabla} \cdot \left(\eta_\psi({\bf r},t) \psi^{1/2}({\bf r},t)\right)$, $\tilde{\eta}_\phi({\bf r},t)={\bf \nabla} \cdot \left(\eta_\phi({\bf r},t) \phi^{1/2}({\bf r},t)\right)$, and $\eta_{\phi,\psi}$ satisfies $<\eta_{\phi,\psi}({\bf r},t)\eta_{\phi,\psi}({\bf r'},t')>=\delta({\bf r}-{\bf r}')\delta(t-t')$. 
 %{\color{blue} Derivation of Eq. \ref{trdensity} and \ref{phi10} can be found in section IB in the Supplementary Information (SI).} 
%The stochastic equations in Eq. \ref{phi10}  contain the same information as the N-body stochastic Langevin equations~\cite{Dean96JPA}. 
%The first terms in both the equations are diffusion.
{{The second term in Eq.(\ref{trdensity}) accounts for the TP-TP interactions (${\bf \nabla }\cdot \left(\psi({\bf r},t)\int_{\bf r'} \psi({\bf r'},t){\bf \nabla}U({\bf r-\bf{r'}})\right)$) and TP-CC interactions, (${\bf \nabla }\cdot \left(\psi({\bf r},t)\int_{\bf r'} \phi({\bf r'},t){\bf \nabla}U({\bf r-\bf{r'}})\right)$).} The influence of the CCs on the TP dynamics is reflected  in the TP-CC coupling. 
%We show below that these two non-linear terms determine the scaling behavior of the dynamical observables for the TPs in both the finite and the long time limit.  Similarly, the second term in Eq.(\ref{phi10}) contains the CC-CC  ($\phi \phi$ fields) and CC-TP interaction ($\phi \psi$) fields. }
The third term in Eq.(\ref{phi10}) results from cell birth and apoptosis, and the fourth term in Eq.(\ref{phi10}) is the non-equilibrium noise. 
Eq.~\ref{trdensity} does not satisfy the  fluctuation-dissipation theorem in the tumor growth phase. %{\color{blue}in the growing phase of tumor}.  

To obtain the exponents describing the dependence of the mean-square displacement (MSD) on time,  we consider a change in scale, $ {\bf r} \rightarrow s {\bf r}$, and $t \rightarrow s^z t$ where $z$ is the dynamic exponent. We use the Parisi-Wu~\cite{Parisi81ES,Himadri06PLA,Himadri06PRE} technique to calculate $z$.  {\color{black} The main features of the method, which are not needed to understand the central results here, are relegated to section I in the SI.} 
 }

{\it{Simulations}}: We simulated a 3D MCS with embedded TPs using an  agent-based model \cite{drasdo2005single, Abdul17Nature, sinha2019spatially}. The cells, treated as interacting soft deformable spheres, grow with time, and divide into two identical daughter cells upon reaching a critical radius ($R_m$). The mean cell cycle time, $\tau = \frac{1}{k_b}$ is 15~hrs is used if not mentioned explicitly. The CCs also undergo apoptosis at the rate $k_a < < k_b$.  The sizes and the number of TPs are held constant.

To determine the TP dynamics, we model the CC-CC, CC-TP, and TP-TP interactions using two potentials (Gaussian and Hertz) to ensure that the qualitative results are robust (simulation details are in section II in the SI). %independent of the nature of the potential (see  SI for details).

  \begin{table}
\begin{tabular}{ c|c|c|c} 
 
   & Theory & Hertz & Gaussian \\ 
 \hline
 $\beta_{TP}$ & 0.57 & 0.12 & 0.11 \\ 
 \hline
 $\alpha_{TP}$ & 2.28 & 2.30  & 2.30\\ 
 \hline
 $\alpha_{CC}$ & 1.45 & 1.47  & 1.50\\ 
 \hline
\end{tabular}
\caption{MSD exponents from theory and simulations.}
\end{table}

The equation of motion for the dynamics of TP and CCs is taken to be \cite{Abdul17Nature}, 
$\dot{\vec{r}}_{i} = \frac{\vec{F}_{i}}{\gamma_i}$,
%\end{equation}
where $\dot{\vec{r}}_{i} $ is the velocity of $i^{th}$ CC or TP, $\vec{F}_{i}$ is the force on $i^{th}$ CC/TP, and $\gamma_i$ is the damping term. 
We used a pressure inhibition mechanism to model the observed growth dynamics in solid tumors \cite{Abdul17Nature}. Dormancy or the growth phase of the CCs  depends on the local microenvironment, determined by the pressure on the $i^{th}$ cell (Figure \ref{tracer_cell}).  Cell division and the placement of the daughter cells add stochasticity in the dynamics of the CCs. We initiated the simulations with 100 TPs and 100 CCs. The spatial coordinates of the CCs and TPs were sampled using a normal distribution with mean zero and standard deviation, $50~\mu m$. 

\floatsetup[figure]{style=plain,subcapbesideposition=top}
\begin{figure}
%\sidesubfloat[]{\includegraphics[width=0.7\linewidth] {tracer_msd_changing_birth.pdf}\label{fig1a}}
%	\par
{\includegraphics[width=1\linewidth] {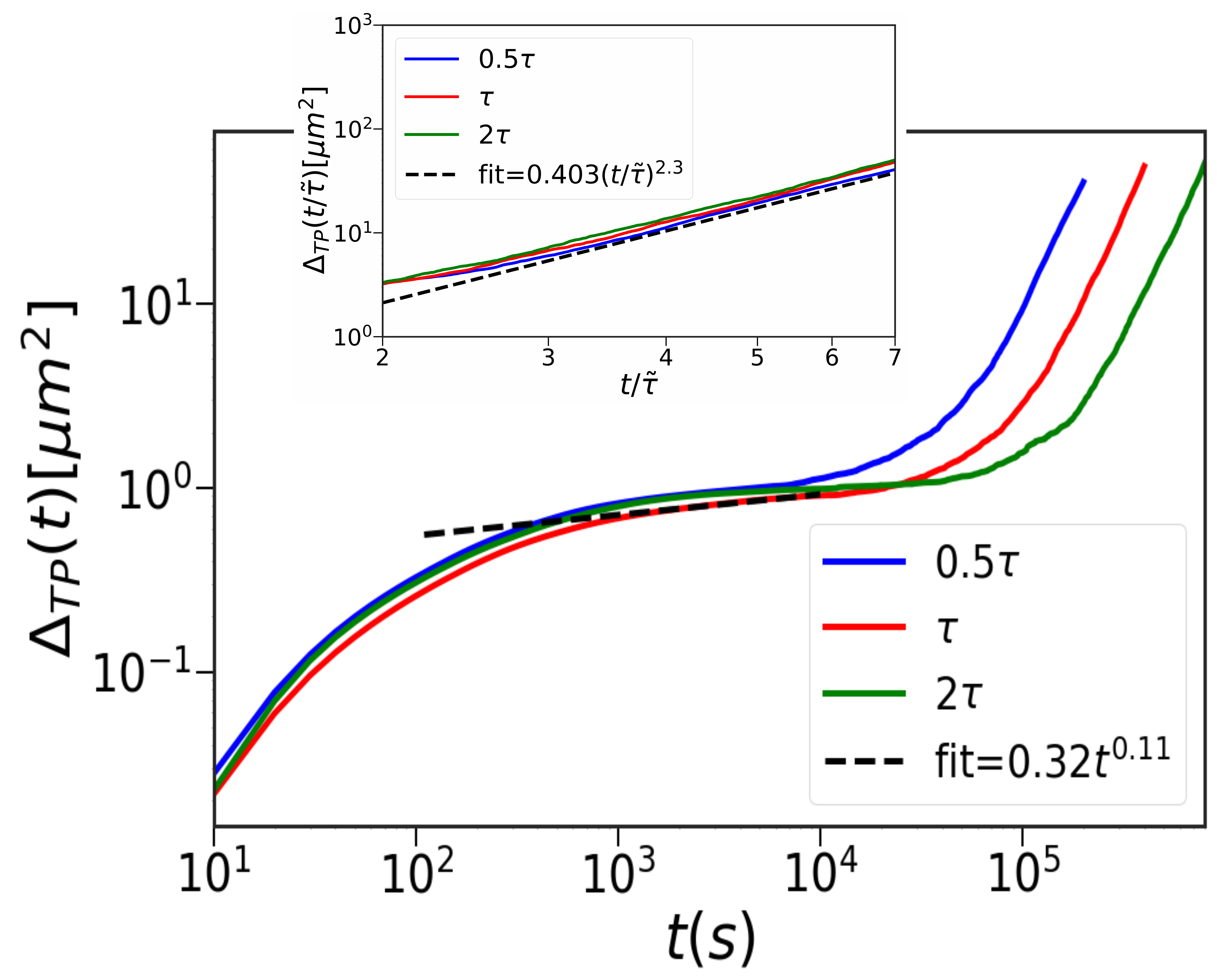}\label{fig1b}} 
\vspace{-.2 in}

%\sidesubfloat[]{\includegraphics[width=0.7\linewidth] {tracer_msd_gaussian_diff_birth.pdf}\label{fig1b}} 
\caption{{\color{black}  MSD of the TPs ($\Delta_{TP}$) using the Gaussian potential. The curves are for  3 cell cycle times (blue ($0.5\tau$), red  ($\tau$), and   green ($2\tau$).  Time  to reach the {hyper-diffusive} behavior, {which is preceded by a jamming regime ($\Delta_{TP}\sim t^{\beta_{TP}}, \beta_{TP}$= 0.11, shown in black dashed line)}, increases with $\tau$.  The inset focusses on the hyper-diffusive regime ($\frac{t}{\tilde{\tau}}>1$). $\tilde{\tau}$ is the cell cycle time for the respective curves. Time is scaled by $\frac{1}{\tilde{\tau}}$. $\Delta_{TP} \sim t^{\alpha_{TP}}, \alpha_{TP}=2.3$ (dashed black line). } }
%The curves correspond to 3 different cell cycle time ( red for $\tau=0.5\tau_{min}$, blue for $\tau=\tau_{min}$ and brown for $\tau=2\tau_{min}$, where $\tau_{min}=54,000 s$). As in the Hertzian scheme, time taken to reach the super-diffusive regime, { preceded by a sub-diffusive regime}, increases on increasing $\tau$. In the long-time ($t>\tau$), we observe super-diffusion for TPs. {The inset shows $\Delta_{TP}$ for the three curves focusing on the super-diffusive regime. The x-axis of the inset plot has been scaled by $\frac{1}{\tau}$ to allow all the curves to overlap in this regime.
\label{tracer_msd_birth}
\end{figure}

%\section{ Results}
{\it {Results:}} {{Theory and simulations predict that, in the limit $t<\frac{1}{k_b}$, the TPs exhibit sub-diffusive behavior (Table I). {\color{black} The non-linear terms 
for the TP-TP and TP-CC  interactions determine the scaling laws.%The non-linear term ${\bf \nabla }\cdot \left(\psi({\bf r},t)\int d{\bf r'} \psi({\bf r'},t){\bf \nabla}U({\bf r-\bf{r'}})\right)$ for the TP-TP interactions contributes to the self-energy term $\Sigma({\bf k}, \omega , \omega_{\tau_f} )$ (in Eq.(\ref{scale1})), and governs the scaling laws.%In the spirit of self-consistent mode coupling theory, we replace $\nu$ by $\Delta \nu$ in the self-energy term $\Sigma({\bf k}, \omega , \omega_{\tau_f} )$ (Eq.(\ref{scale1})). 
~A power counting  analysis shows that  the dynamical exponent $z=2+\frac{d}{2}$.}
%Using Eq. (\ref{scale1}), we obtain $k^{z}\sim k^{d-z+4}$, which leads to $z=2+\frac{d}{2}$.
As a consequence, the MSD of a TP behaves as, 
$\Delta(t)_{TP} = <[r(t)-r(0)]^2>\sim t^{2/z} = t^{\beta_{TP}}$. Because $\beta_{TP}$ is less than unity we surmise that the TPs are jammed.
%In 3D, we find that $\beta_{TP}^{T}=\frac{4}{7}= 0.57$. Thus, the TPs undergo sub-diffusive or jamming motion.
%and $\alpha=\frac{4}{8}= 0.5$ for interaction potential $U_1=U_0/\cosh^2(r/a)$, implying sub-diffusive behavior. %Although sub-diffusive behavior is preserved at intermediate times, 
%The scaling exponent $\beta_{TP}^{T}$ depends (section IIIA in the SI)  on the form of the interaction potential. % implying that the behavior of MSD in the intermediate time limit is non-universal. 
%For example, if the TP-TP interaction is modeled as $U_1=U_0/\cosh^2(r/a)$ instead of a Gaussian (Eq.(\ref{potential})), we obtain $\alpha=\frac{4}{8}= 0.5$, implying sub-diffusive behavior is preserved at intermediate times (See the discussion in Appendix F). 
%The sub-diffusive behavior is a consequence of {caging of TP's by surrounding CCs and TPs}, which is vividly illustrated in our simulations.

We also calculated $\Delta_{TP}(t)=\langle[{\bf r}(t)-{\bf r}(0)]^2\rangle$, by averaging over $\approx 2,000$ trajectories, for the Gaussian (Figure \ref{tracer_msd_birth}) and Hertz (Figure {\color{blue}S2a in the SI}) potentials.  Similar behavior is found in both cases. The duration of the plateau ({Figure \ref{tracer_msd_birth}) increases as the cell cycle time increases. Although  jamming behavior predicted theoretically is consistent with the simulations the $\beta_{TP}$ exponents differ (Table I). %at short times, but  $\beta_{TP}^{S}$ is less than $\beta_{TP}^{T}$. 

%{The length of the plateau in the intermediate time increases as the cell cycle time ($\tau$)} increases. %The caging effect is stronger as the cell cycle time increases.

\floatsetup[figure]{style=plain,subcapbesideposition=top}
\begin{figure}
\includegraphics[width=1\linewidth] {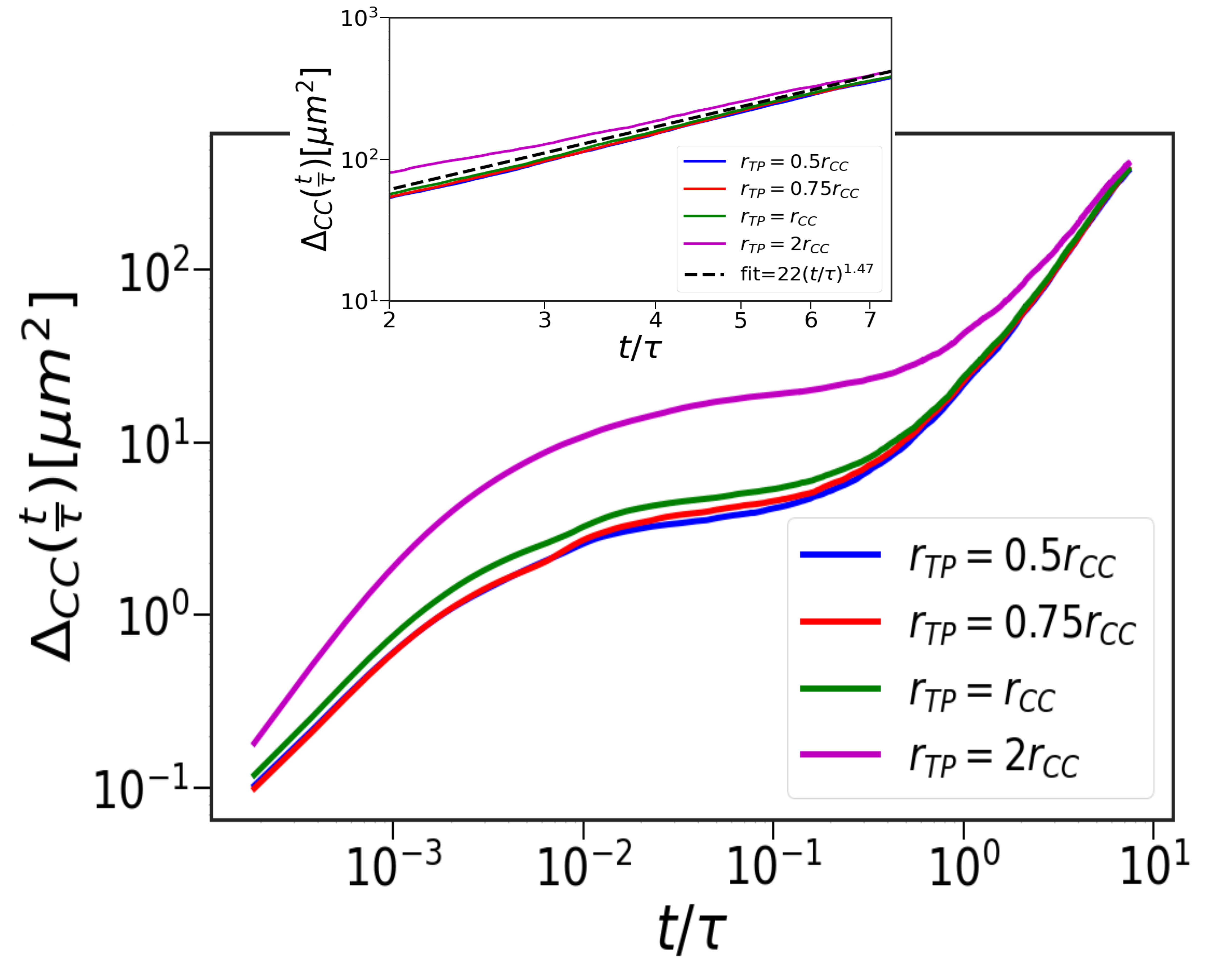}
\vspace{-.3 in}
\caption{{\color{black} CC dynamics in the presence of the TPs. $\Delta_{CC}$ using the Hertz potential. The curves are for different TP radius ( magenta - $r_{TP}=$ { 2} $r_{CC}$, green - $r_{TP} = r_{CC}$, and red - $r_{TP} = 0.75r_{CC}$ , and blue - $r_{TP} = 0.5r_{CC}$, where $r_{CC} =4.5 \mu m$ is the average cell radius. Inset shows  $\Delta_{CC}$ for the four curves, focusing on the super-diffusive regime.  The black dashed line is drawn with  $\alpha_{CC}=1.47$.}}
\label{fig4aa}
\end{figure}

In the $t>\frac{1}{k_b}$ limit, theory and simulations predict hyper-diffusive dynamics, $\Delta_{TP} \sim t^{\alpha_{TP}}$ with  $\alpha_{TP}>2$. {\color{black}Non-linearities, due to the TP-CC interactions together with cell division and apoptosis,  determine the long-time behavior of the TPs. Using the scaling analysis as before, the  theory predicts hyper-diffusive dynamics for the TPs (Table I). Variations in $k_b$ do not change the value of $\alpha_{TP}$. It merely changes the coefficients of the linear term and the value of $\frac{2k_b}{k_a}$. Therefore, $\alpha_{TP}$  is independent of the cell cycle time in the long time limit.}
Simulation results are in excellent agreement with the theoretical predictions. For times exceeding the cell cycle time, ($t>\tau$), the TPs exhibit hyper-diffusive behavior (Figure \ref{tracer_msd_birth}). For the Gaussian potential, we obtain $\alpha_{TP} \approx$  2.3  ( see the inset in Figure \ref{tracer_msd_birth}) for $0.5\tau$, $\tau$ and $2\tau$. Thus, $\alpha_{TP}$ is independent of cell cycle time. 
{\color{black} Note that TP-TP interactions play an insignificant role in the dynamics of TPs or the CCs (Figure S3 in the SI). They merely alter the amplitude of $\Delta_{TP}$ in the intermediate time. They do not affect  the long time dynamics}

\floatsetup[figure]{style=plain,subcapbesideposition=top}
\begin{figure}
%\sidesubfloat[]{\includegraphics[width=0.8\linewidth] {tracer_scat_diff_birth.eps}\label{fig2a}}
%	\par
{\includegraphics[width=1\linewidth] {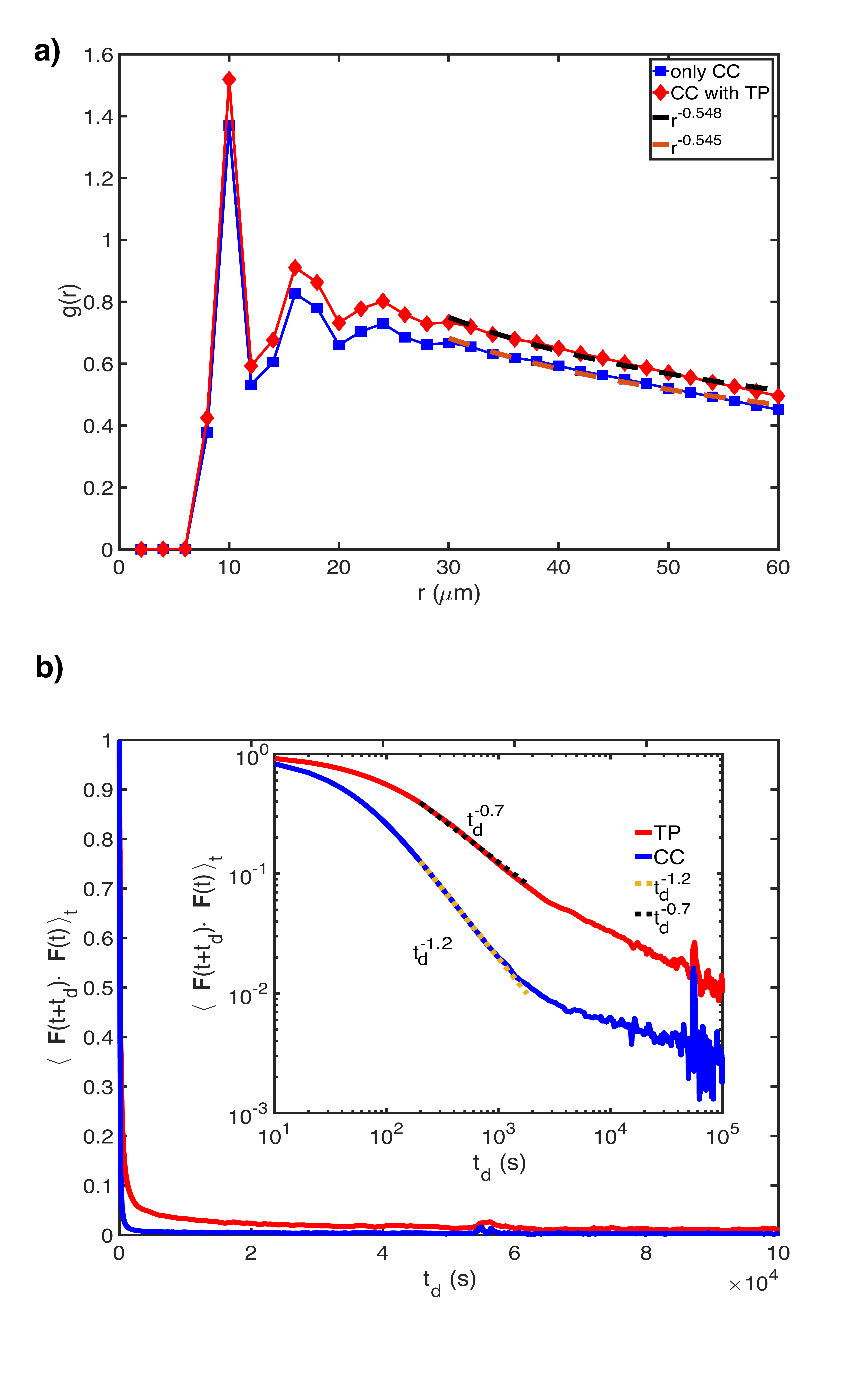}} 
\vspace{-.6 in}

%\sidesubfloat[]{\includegraphics[width=0.8\linewidth] {tracer_scat_gaussian_diff_birth.eps}\label{fig2b}} 
\caption{ {\color{black} {\bf (a)} CC correlation function, ($g(r)$), as a function of inter-cellular distance. Red (blue) curve shows $g(r)$ in presence (absence) of TPs. The two dashed lines (black and orange) are power law fits to $g(r)$ in the large $r$ limit. {\bf (b)} Force force autocorrelation (FFA), as a function of delay time ($t_d$). The red (blue) curve shows the FFA for TPs (CCs). Inset shows FFA on log-log scale. The black (yellow) dashed line is a  power law fit with exponent of -0.7 (-1.2). } }
\label{gr_ffa}
\end{figure}

Figure \ref{fig4aa} shows that changing the tracer size affects only the amplitude of $\Delta_{CC}$ in the intermediate time without altering the $\alpha_{TP}$ values (see also figures S4 and S5 in the SI). This is because the TP size only changes the nature of the short-range interactions without introducing any new scale.  Since, $\alpha_{CC}$ is a consequence of the long-range spatial and temporal correlations that emerge because of the inequality $k_b >> k_a$, it is independent of the tracer size (section IIIE of SI).

{\color{black}In the absence of the TPs,  birth and apoptosis determine the CC dynamics in the long time regime, yielding $\alpha_{CC}=1.33$ \cite{Abdul17Nature,Himadri18PRE}. 
When TPs are present, the CCs continues to exhibit super-diffusive motion with a modest increase in $\alpha_{CC}$ (Table I).    The CC pair-correlation, $g(r)=\frac{V}{4\pi r^2 N^2} \sum_{i=1}^{N}\sum_{j\neq i}^{N}\delta(r-|{\bf r}_i- {\bf r}_j|) \sim r^{-0.5}$, in the presence and absence of TPs at $t\approx 8 \tau$  (Figure \ref{gr_ffa}a).  The dynamically-induced CC correlations is independent of the TPs, thus explaining the insignificant effect of the  TPs on  the CC dynamics. }

{\color{black} To explain the finding, $\alpha_{TP}>\alpha_{CC}$, we calculated the force-force autocorrelation function, FFA=$\langle {\bf F}(t+t_d)\cdot {\bf F}(t)\rangle_t$. Here, $\langle ... \rangle_t$ is the time average and $t_d$ is a delay time. Since, the TPs (CCs) exhibit hyper-diffusion (super-diffusion), we expect that the FFA of TPs should decay slower relative to the CCs, which is confirmed in  Figure \ref{gr_ffa}b, which  shows that the FFA for the TPs (CCs) decays as $t_d^{-0.7}$ ($t_d^{-1.2}$).  Thus, the TP motion is significantly more persistent than the CCs, which explains the hyper-diffusive nature of the TPs.}

\floatsetup[figure]{style=plain,subcapbesideposition=top}
\begin{figure}
%\sidesubfloat[]{\includegraphics[width=0.8\linewidth] {tracer_scat_diff_birth.eps}\label{fig2a}}
%	\par
{\includegraphics[width=1\linewidth] {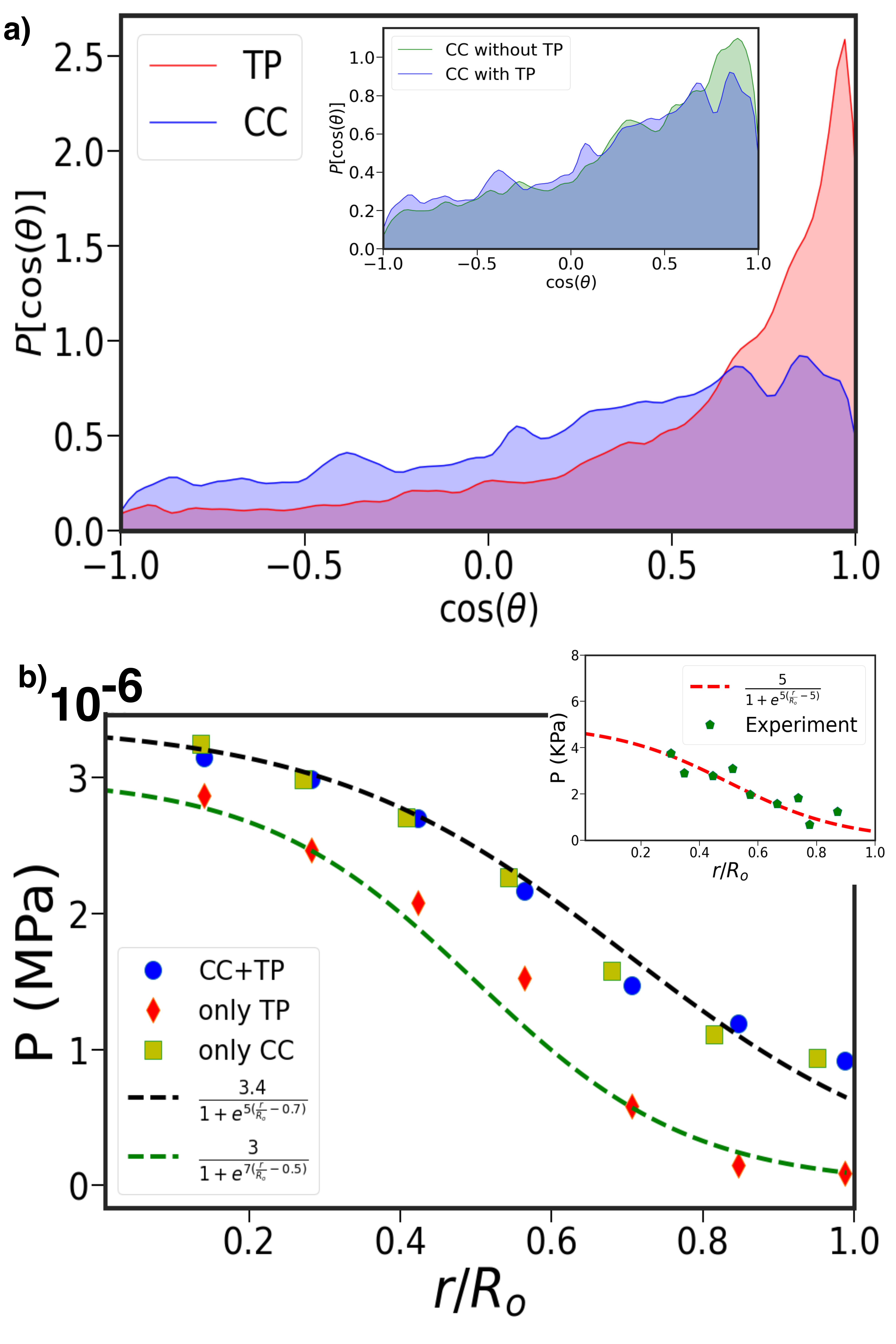}} 
\vspace{-.2 in}
%\sidesubfloat[]{\includegraphics[width=0.8\linewidth] {tracer_scat_gaussian_diff_birth.eps}\label{fig2b}} 
\caption{ {\bf (a)} Distribution of $\cos(\theta)$ for the TPs and CCs.  $\theta$ is the angle between two consecutive steps along a  trajectory. The red (blue) plot is for the TPs (CCs). The distribution is skewed to $\cos(\theta)>0$, indicative of persistent motion. Extent of skewness is greater for TPs than CCs. The inset shows the $P[\cos (\theta)]$ of CCs with and without TPs. {\bf (b)}{\color{black}  Pressure as a function of the radial distance ($r$ scaled by $R_0$: tumor radius $\approx 100 \mu m$) from tumor center. The blue circles correspond to local pressure measured using both CCs and TPs. The red data points is local pressure measured using just the TPs using simulations containing both the CCs and TPs. The greenish yellow squares give the local pressure obtained in simulations  with only the CCs. The black and green dashed lines is logistic fit. The inset shows the radial pressure profile in experiments (green data points). The red dashed lines are logistic fits. Note that the magnitude of pressure measured in simulations and experiments are in qualitative agreement.} }
\label{theta_si}
\end{figure}

A mechanistic explanation for $\alpha_{TP}>\alpha_{CC}$ can be gleaned  from the simulations. To illustrate the difference between the dynamics of the TPs and CCs, we calculated the angle ($\theta$) between two consecutive time steps in a trajectory. We define $\cos(\theta(t,\delta t))=\frac{\delta {\bf r}(t+\delta t) \cdot \delta {\bf r}(t)}{|\delta {\bf r}(t+\delta t)||\delta {\bf r}(t))|}$, where $\delta {\bf r}(t)={\bf r}(t+\delta t)-{\bf r}(t)$. Figure \ref{theta_si}a shows the ensemble and time averaged $\cos \theta$ distribution for $\frac{\delta t}{\tau} = 1$. If the motion of the CCs and TPs were diffusive, the distribution of $\cos \theta$ would be uniformly distributed from -1 to 1. However, Figure \ref{theta_si}a shows that $P(\cos\theta)$ is skewed towards unity implying that the motion of both the TPs and CCs are persistent. Interestingly, the skewness is more pronounced for the TPs compared to the CCs (Figure \ref{theta_si}a). During every cell division, the motion of CCs is randomized and hence the persistence is small compared to TPs. This explains the hyper-diffusive (super-diffusive) for the TPs (CCs) and thus $\alpha_{TP}>\alpha_{CC}$ in the long time limit (see Figure S6 which shows TPs move much more straighter than CCs). 

{\color{black} To assess the effectiveness of the TPs as sensors of the local microenvironment, we calculated the radial pressure (Eq S28 in the SI) profile at $t=7.5\tau$, which has been measured \cite{Dolega17NC}.  We measured pressure profiles using Eq. S28 (in the SI) for the system with both CCs and TPs.  We also calculated pressure experienced  just by the TPs. Finally, we computed pressure in a system consisting of solely  the CCs.  Two pertinent remarks  about  Figure (\ref{theta_si}b are worth making. (i) All three curves  qualitatively capture the pressure profile found in the experiments. Pressure decreases roughly by factor of four, as the distance $r$  from the center of the tumor increases. The pressure is almost constant in the core, with a decrease that can be fit using the logistic function, as the boundary of the tumor is reached  (Figure (\ref{theta_si}b). The high core pressure is due to small number of cell divisions. As a consequence, the CCs are jammed, leading to high internal pressure.  As $r$ increases, the CCs  proliferate, resulting in a decrease in self generated stress, and consequently a decreases in the pressure (Figure (\ref{theta_si}b).(ii) Most importantly, the CC pressure profiles are unaltered even in the presence of the TPs (compare blue circles and greenish yellow squares in Figure (\ref{theta_si}b), which shows that the latter does faithfully report the microenvironment of the CCs.}

{\color{black} We conclude with a few comments.   (i) The excellent agreement for $\alpha_{TP}$ between simulations and theory suggests that  the CC dynamics is determined by the overall tumor growth, and not the details of the TP-CC interactions. Although $\alpha_{TP} > \alpha_{CC}$,  the CCs move a larger distance than the TPs (compare Figure \ref{tracer_msd_birth} and Figure  \ref{fig4aa}), which implies that the effective diffusion constant of the CCs is greater than the TPs.  (ii) The radial dependence of the pressure (Figure \ref{theta_si}b), with the core experiencing higher values than the periphery cells, is another manifestation of non-equilibrium that is induced by an imbalance in division and apoptosis rates of the CCs. (iii) Most importantly, we find that the long time dynamics of CCs and the pressure profiles are unchanged when the TPs are embedded in the MCS. Therefore, it is justified to use the TPs as tumor stress sensors.}  
%The pressure profile decreases monotonically with the radius of the growing spheroid. The higher pressure in the core indicates fewer cell-division. The pressure decreases exponentially with the radius showing the gradual increase of the proliferation of CCs.	

\textbf{Acknowledgements}
 \noindent 
We thank Mauro Mugnai, Hung Nguyen, Rytota Takaki and Davin Jeong for useful discussions and comments on the manuscript.  This work is supported by the National Science Foundation (PHY 17-08128), and the Collie-Welch Chair through the Welch Foundation (F-0019).

\providecommand*{\mcitethebibliography}{\thebibliography}
\csname @ifundefined\endcsname{endmcitethebibliography}
{\let\endmcitethebibliography\endthebibliography}{}

%\bibliographystyle{rsc}
%\bibliography{tracer.bib}
\end{document}

\title{Supplementary Information: Far from equilibrium dynamics of tracer particles embedded in a growing multicellular spheroid }
\author{Himadri S. Samanta}\affiliation{Department of Chemistry, University of Texas at Austin, TX 78712}
\author{Sumit Sinha}\affiliation{Department of Physics, University of Texas at Austin, TX 78712}
\author{D. Thirumalai}\affiliation{Department of Chemistry, University of Texas at Austin, TX 78712}

\date{\today}
%\begin{abstract}
 % \end{abstract}

\maketitle
\clearpage
%\tableofcontents
%\def\s{\rule{0in}{0.28in}}

%\section{Theory}

\section{Theory}

\subsection{Time-dependent equations for TP and CC densities}
\begin{figure}
\subfloat[]{\includegraphics[width=0.55\linewidth] {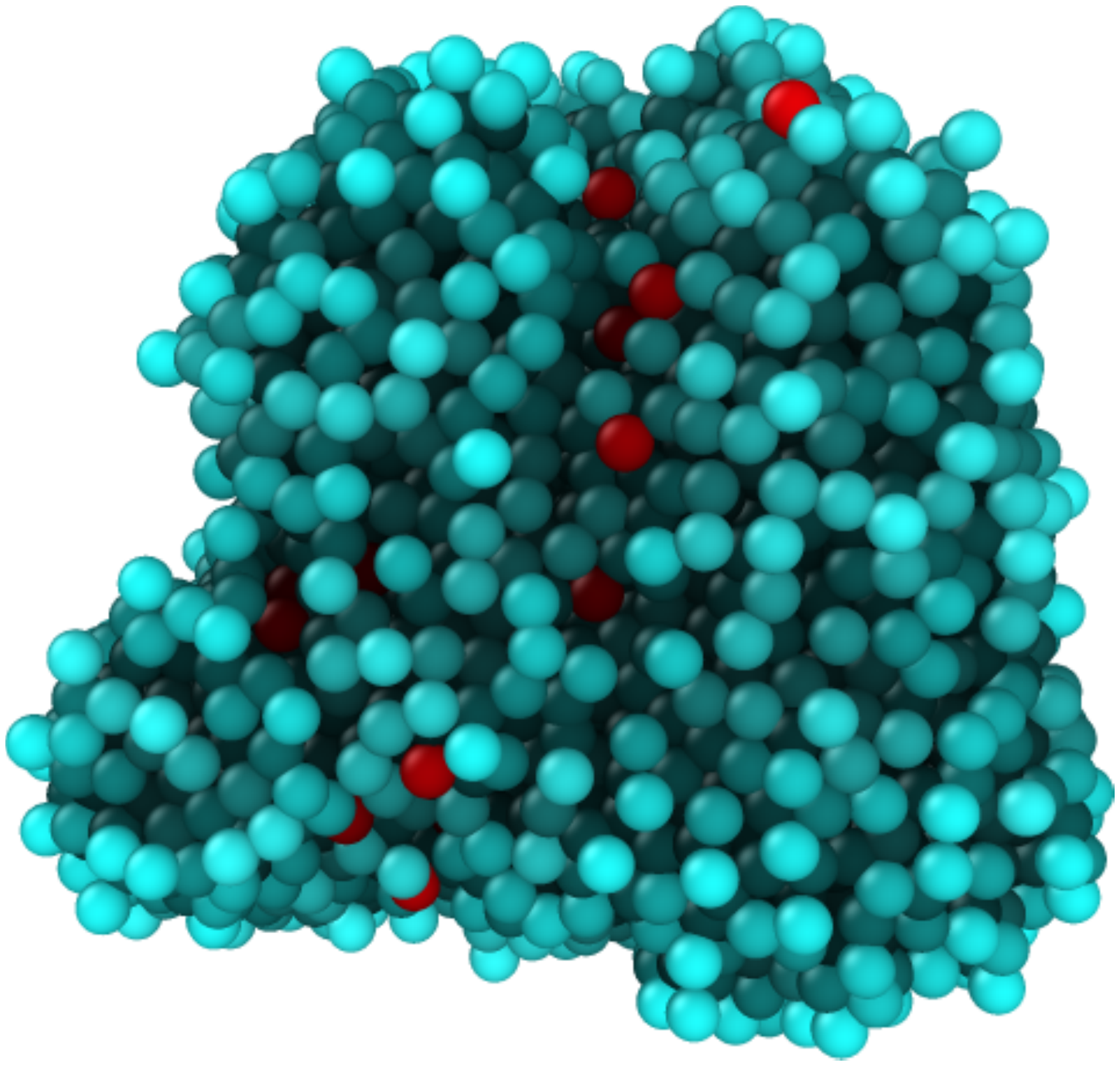}\label{tumor_pic1}}
%\par
\subfloat[]{\includegraphics[width=0.55\linewidth] {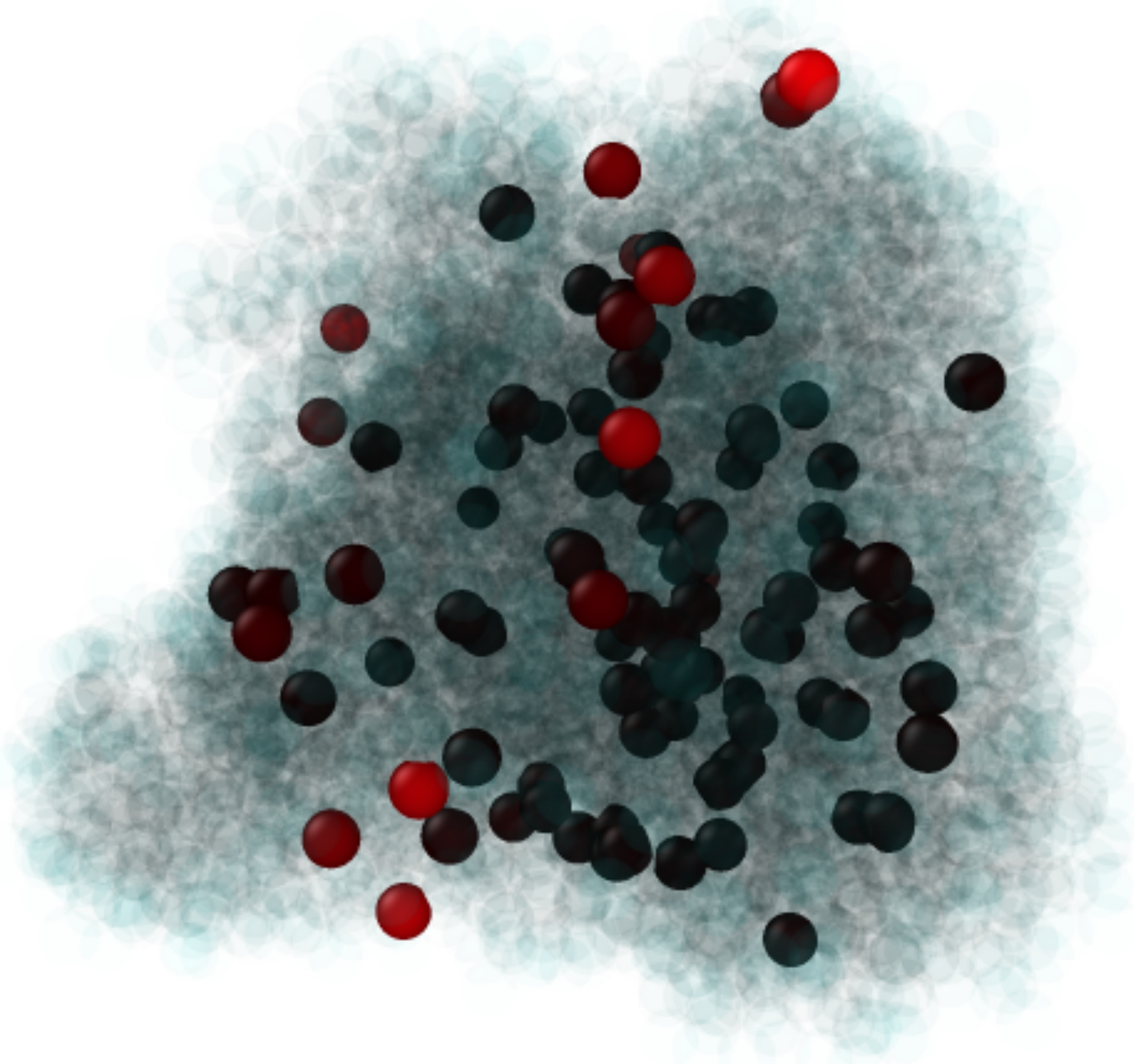}\label{tumor_pic2}} 

%\sidesubfloat[]{\includegraphics[width=0.65\linewidth] {tumor_pic2.pdf}\label{tumor_pic2}} 
\caption{Snapshot of tumor simulation with embedded tracers. {\bf (a)} A 3D simulated spheroid consisting of approximately 4800 CCs and 100 TPs. The CCs are in cyan, and the tracers are in red.  {\bf (b)} The spheroid shown above was rendered by making the CC cells transparent (light colored cyan) in order to show the interior of the spheroid. The TPs are opaque. Some of the TPs appear black because it is a depiction of a 3D image. The purpose of displaying these snapshots is to visually show that the TPs are randomly distributed within the multicellular spheroid, implying that their migration is largely determined by the forces arising from the CCs.}
\label{tumor_pic}
\end{figure}

Let us consider the dynamics of the tracer particles (TPs) in a growing tumor spheroid. The TPs experience systematic short-range interactions due to volume excluded from the neighboring TPs and the cancer cells (CCs). In addition, they are also subject to a random force characterized by a Gaussian white noise spectrum.
For mathematical convenience, the inter-cell interactions are modeled as a sum of  attractive, and repulsive excluded volume interactions. We assume that the dynamics of the system, consisting of the CCs and TPs ({see Figure \ref{tumor_pic} for snapshots generated in simulations}) can be described by {the  overdamped} Langevin equation,
\begin{equation}
\frac{ d{\bf r}_i}{ dt}=-\sum_{j=1}^N \nabla U (|{\bf r}_i - {\bf r}_j|) +{\boldsymbol \eta}_i (t),
\label{eqn_of_motion}
\end{equation}
where $ {\bf r}_i$ is the position of a CC or a TP, and ${\boldsymbol \eta}_i(t)$ is a Gaussian random force with white noise spectrum. {To keep the problem theoretically tractable, the form of $U(|{\bf r}_i-{\bf r}_j|)$ between a pair of particles (can be either TP-TP, TP-CC or CC-CC) is taken to be }, 
\begin{equation}\label{potential}
U(|{\bf r}(i)-{\bf r}(j)|)=\frac{\nu}{(2\pi\lambda^2)^{3/2}}e^{\frac{-|{\bf r}(i)-{\bf r}(j)|^2}{2\lambda^2}}  
 -\frac{\kappa}{(2\pi\sigma^2)^{3/2}}e^{\frac{-|{\bf r}(i)-{\bf r}(j)|^2}{2\sigma^2}},
\end{equation}
where, $\lambda$ and $\sigma$ are the ranges of the repulsive and attractive interactions, and $\nu$ and $\kappa$ are the interaction strengths. Thus, the interactions involving the mixture of CCs and TPs are identical.

{When equation \ref{eqn_of_motion} is used to characterize the dynamics of the TPs}, the potential $U_{TP}$ contains both the TP-TP and TP-CC interactions with the corresponding attractive {(repulsive) interaction ranges being $\sigma_1 (\lambda_1)$ and $\sigma_2 (\lambda_2)$,  respectively.}
The potential $U_{CC}$ for the CCs mimics cell-cell adhesion (second term in the above equation) and excluded volume interactions, and the CC-TP interactions.  

%In terms of the density function for a single cell $\phi_i({\bf r},t)=\delta[\bf r-{\bf r}_i(t)]$, a 

A closed form Langevin equation for the CC density, {$\phi({\bf r},t)=\sum_i \phi_i({\bf r},t)$}, where $\phi_i({\bf r},t)=\delta[\bf r-{\bf r}_i(t)]$, can be obtained using the method introduced elsewhere~\cite{Dean96JPA}. The time evolution of $\phi({\bf r},t)$ is given by,
\begin{eqnarray}\label{phicell}
&&\frac{\partial \phi({\bf r},t)}{\partial t}=  {\bf \nabla }\cdot \left(\phi({\bf r},t)\int_{\bf r'} \left[\psi({\bf r'},t){\bf \nabla}U_{CC-TP}({\bf r-\bf{r'}})+ \phi({\bf r'},t){\bf \nabla}U_{CC}({\bf r-\bf{r'}})\right]\right)\\ \nonumber &&+D_{\phi} \nabla^2 \phi({\bf r},t)+{\bf \nabla} \cdot \left(\eta_{\phi}({\bf r},t) \phi^{1/2}({\bf r},t)\right),
\end{eqnarray}
where $\eta_\phi$ satisfies $<\eta_\phi({\bf r},t)\eta_\phi({\bf r'},t')>=2D_{\phi}\delta({\bf r}-{\bf r}')\delta(t-t')$.
Similarly,
the evolution of the density function for {a single} TP, $\psi({\bf r},t)=\sum_i \psi_i({\bf r},t)=\sum_i\delta[\bf r-{\bf r}_i(t)]$, may be written as,
\begin{widetext}
 \vspace{-.2 in}
\begin{eqnarray}\label{trdensity}
\frac{\partial \psi({\bf r},t)}{\partial t}&=&D_\psi \nabla^2 \psi({\bf r},t)+ {\bf \nabla }\cdot \left(\psi({\bf r},t)\int_{\bf r'} [\psi({\bf r'},t){\bf \nabla}U_{TP}({\bf r-\bf{r'}})
\right.\\ \nonumber &&
+ \phi({\bf r'},t){\bf \nabla}U_{TP-CC}({\bf r-\bf{r'}})]\left.\right)+{\bf \nabla} \cdot \left(\eta_\psi({\bf r},t) \psi^{1/2}({\bf r},t)\right).
\end{eqnarray}
\end{widetext}
 \vspace{-.2 in}
 where $\eta_\psi$ satisfies $<\eta_\psi({\bf r},t)\eta_\psi({\bf r'},t')>=2D_{\psi}\delta({\bf r}-{\bf r}')\delta(t-t')$.
%{The second term in Eq.(\ref{trdensity}) accounts for the TP-TP/TP-CC interactions.}  The influence of CCs on the TP dynamics arises explicitly through the {second} term in Eq.(\ref{trdensity}). We show below that these non-linear terms determine the scaling behavior of the {dynamical observables for the TPs in both finite and long time limit}.

We modify the density evolution for the CCs phenomenologically by adding a source term describing cell division and apoptosis, and a noise term that breaks the CC number conservation. {\color{black}These terms can be formally derived as follows.  
Birth and apoptosis reactions are given by
$X \xrightarrow[]{k_b} X+X$ and $X+X\xrightarrow[]{k_a/\Omega} X$, where $k_a/\Omega$ is the apoptosis rate of distinct pairs of cells, and $\Omega $, the volume, will eventually be set to infinity.
Let the $\rho $ be the density of cells. 
The master equation for the evolution of the probability $P_n (t)$ is given by,
\begin{equation}\label {master}
\frac{dP_n(t)}{dt}=k_b (n-1) P_{n-1} -k_b n P_n - \frac{k_a}{\Omega} \frac{n(n-1)}{2} P_n + \frac{k_a}{\Omega} \frac{n(n+1)}{2} P_{n+1}
\end{equation}

\noindent Eq.(\ref{master}) can be expanded up to second order in $n$ and the resulting equation is given by
\begin{eqnarray}\label {master1}
\frac{dP_n(t)}{dt}&=&k_b (1-\frac{\partial}{\partial n}+\frac{1}{2}\frac {\partial^2}{\partial n^2}) (n) P_{n} -k_b n P_n \\ \nonumber&&- \frac{k_a}{\Omega} \frac{n(n-1)}{2} P_n + \frac{k_a}{\Omega} (1-\frac{\partial}{\partial n}+\frac{1}{2}\frac {\partial^2}{\partial n^2})\frac{n(n-1)}{2} P_{n}\\ \nonumber
&=& \frac{\partial }{\partial n} [-k_b n + \frac{k_a}{2\Omega}n(n-1)] P_n (t) +
\frac{\partial^2}{\partial n^2} [\frac{k_b}{2} n + \frac{k_a}{4\Omega}n(n-1)] P_n (t)
\end{eqnarray}
The corresponding Langevin equation for the density $\rho=n/\Omega$ is, 
\begin{equation}
\frac{\partial \rho (t)}{\partial t} = k_b \rho(t)-\frac{k_a}{2} \rho(t) (\rho(t)-\frac{1}{\Omega}) +\frac{1}{\sqrt{2\Omega}}\sqrt{{k_b} \rho(t)+\frac{k_a}{2} \rho(t) (\rho(t)-\frac{1}{\Omega})}f(t)
\end{equation}

In order to account for spatial variations, we generalize this scheme by considering the concentration in the $i$th volume element, 
 \begin{equation}
\frac{\partial \rho_i (t)}{\partial t} = k_b \rho_i(t)-\frac{k_a}{2} \rho_i(t) (\rho_i(t)-\frac{1}{\Omega}) +\frac{1}{\sqrt{2\Omega}}\sqrt{{k_b} \rho_i(t)+\frac{k_a}{2} \rho_i(t) (\rho_i(t)-\frac{1}{\Omega})}f_i(t),
\end{equation}
where $<f_i(t)f_j(t')>=2\delta_{ij}\delta(t-t')$. In the continuum limit, $\rho_i \rightarrow \rho({\bf r},t)$, $f_i(t)\rightarrow f({\bf r},t)$, and $\delta_{ij}\rightarrow \Omega \delta({\bf r}-{\bf r'})$, we get the following equation,
\begin{eqnarray}
\frac{\partial \rho ({\bf r},t)}{\partial t} &=& (k_b +\frac{k_a}{2\Omega}) \rho({\bf r},t)-\frac{k_a}{2} \rho({\bf r},t)^2 +\sqrt{ (k_b -\frac{k_a}{2\Omega}) \rho({\bf r},t)+\frac{k_a}{2} \rho({\bf r},t)^2 }f({\bf r},t) \\ \nonumber
&=& \frac{k_a}{2}  \rho({\bf r},t)( (\frac{2k_b}{k_a} +\frac{1}{\Omega})-\rho({\bf r},t))+\sqrt{ (k_b -\frac{k_a}{2\Omega}) \rho({\bf r},t)+\frac{k_a}{2} \rho({\bf r},t)^2 }f({\bf r},t),
\end{eqnarray}
$<f({\bf r},t)f({\bf r'},t')>=\delta({\bf r}-{\bf r'})\delta (t-t')$.

 By letting $\Omega \rightarrow \infty$, the time evolution of the density $\rho$ becomes,
 \begin{equation}
\frac{\partial \rho ({\bf r},t)}{\partial t}= \frac{k_a}{2}  \rho({\bf r},t)\left( \frac{2k_b}{k_a} -\rho({\bf r},t)\right)+\sqrt{ k_b  \rho({\bf r},t)+\frac{k_a}{2} \rho({\bf r},t)^2 }f({\bf r},t).
\end{equation}
We add the terms on the right hand side in the density equation for the CCs.}

The final Langevin equation, for the time-dependent changes in $\phi({\bf r},t)$ is~\cite{Himadri18PRE},
\begin{widetext}
% \vspace{-.2 in}
\begin{eqnarray}
\label{phi10}
\frac{\partial \phi({\bf r},t)}{\partial t}&=& D_\phi \nabla^2 \phi({\bf r},t)+ {\bf \nabla }\cdot \left(\phi({\bf r},t)\int d{\bf r'}  [\psi({\bf r'},t){\bf \nabla}U_{CC-TP}({\bf r-\bf{r'}})
\right. \\ \nonumber && \left.
+ \phi({\bf r'},t){\bf \nabla}U_{CC}({\bf r-\bf{r'}})]\right)
+ \frac{k_a}{2} \phi(\frac{2k_b}{k_a}-\phi)+{\bf \nabla} \cdot \left(\eta_\phi({\bf r},t) \phi^{1/2}({\bf r},t)\right)
+\sqrt{k_b \phi+\frac{k_a}{2} \phi^2} f_\phi \, ,
\end{eqnarray}
\end{widetext}
 %\vspace{-.5 in}
where $f_\phi$ satisfies $<f_\phi({\bf r},t)f_\phi({\bf r'},t')>=\delta({\bf r}-{\bf r}')\delta(t-t')$.
The source term $\propto \phi(\phi_0-\phi)$, represents the cell division and apoptosis, with $\phi_0=\frac{2k_b}{k_a}$\cite{Doering03PA,Gelimson15PRL}.~The coefficient of $f_\phi$, given by~$\sqrt{k_b \phi+\frac{k_a}{2}\phi^2} $, is the strength of the noise corresponding to number fluctuations of the CCs, and is a function of the CC density. %Because we are interested 
%dynamics of the CCs and TPs in the finite and  long-time limit we use the Parisi-Wu technique~\cite{Parisi81ES,Himadri06PLA,Himadri06PRE}. 

 %A major difficulty in the TP dynamics coupled with CCs arises due to the breakdown of fluctuation-dissipation theorem (FDT) in Eq. (\ref{phi10}). Diagrammatic expansions for the response function and the correlation functions could be used to solve Eq.\ref{phi10}. Because we are interested dynamics of the CCs and TPs in the finite and  long-time limit we use the Parisi-Wu technique~\cite{Parisi81ES,Himadri06PLA,Himadri06PRE}. 

{\color{black}To simplify the multiplicative noise terms (last term in Eq. (\ref{trdensity}) and the last two terms in Eq. (\ref{phi10})), we assume that the density fluctuates around a constant value. We write $\Psi({\bf r},t) = \Psi_0 + \Psi_1({\bf r},t)$ ($\Psi({\bf r},t)$ can be either $\phi$ or $\psi$), and expand Eq. (\ref{trdensity}) and Eq. (\ref{phi10}) in terms of $\Psi_1$ up to the lowest order in nonlinearity.% We are interested how the interplay of adhesion interactions and birth-apoptosis processes determine the TP density fluctuations $ \psi_1$ and the CC density fluctuations $ \phi_1$ at time scale of cell cycle times.
~We consider the scaling behavior under a change of scale given by ${\bf{r}} \rightarrow s\bf{r}$ and $t \rightarrow s^z t$ with $z $ being the dynamic exponent.

%To anticipate the consequence of non-linearity in the finite ($t<\frac{1}{k_b}$) and long ($t>\frac{1}{k_b}$) time limit}, we introduce a change of scale $\psi_1\rightarrow s^{\chi_1} \psi_1 $, and $\phi_1 \rightarrow s^{\chi_2} \phi_1$, where $\chi_1 $ and $\chi_2$ are the exponents corresponding to TP and CC density fluctuations respectively. 
The interplay between adhesion interactions and stochastic birth-apoptosis processes gives rise to long-range correlations in the density fluctuations of the TPs and CCs.
We consider the scaling of the correlation of density fluctuations for TP as $< \psi_1 ({\bf{r'}},t) \psi_1 ({\bf{r}},t)> \sim |\bf{r'}-\bf{r}|^{2\chi_1}$, and similarly for CC, $< \phi_1 ({\bf{r'},t)}\phi_1 ({\bf{r}},t)> \sim |\bf{r'}-\bf{r}|^{2\chi_2}$, where $\chi_1 $ and $\chi_2$ are the exponents corresponding to TP and CC density fluctuations respectively. }
For $t \lesssim {k_b}^{-1}$, 
the scaling of the dynamical observables for the TPs is governed by the TP-TP and TP-CC interactions. In Fourier space, the non-linear term ($q.(k-q)\psi_1(q)\psi_1(k-q)$) for the TP-TP interactions scales (Eq.~(\ref{trdensity})) as $q^{2-2\chi_1}$. Similarly, the non-linear term ($q.(k-q)\phi_1(q)\psi_1(k-q)$) for the TP-CC interactions scales as $q^{2-\chi_1-\chi_2}$. The degree of non-linearity for both the TP-TP and TP-CC interactions is $q^{2-2\chi_1}$, by noting that $\chi_1 \sim \chi_2$. Therefore, both TP-CC and TP-TP interactions exhibit the same scaling for the dynamical properties of TPs in the $t \lesssim {k_b}^{-1}$ limit. 
At long times ($t>\frac{1}{k_b}$), the birth-apoptosis term $\phi_1(q)\phi_1(k-q)$ (scales as $q^{-2\chi_2}$), dominates over the short-range CC-CC interactions $q.(k-q)\phi_1(q)\phi_1(k-q)$ ( scales as $q^{2-2\chi_2}$). Therefore, when $t>\frac{1}{k_b}$, the birth-apoptosis non-linearity determines the scaling properties for the TPs through the TP-CC interactions.
%the scaling of dynamical observables for TPs is governed by the non-linear term ${\bf \nabla }\cdot \left(\psi({\bf r},t)\int d{\bf r'} \psi({\bf r'},t){\bf \nabla}U({\bf r-\bf{r'}})\right)$ which describes TP-TP interactions. In the long time limit, TP-CC non-linear interaction term ${\bf \nabla }\cdot \left(\psi({\bf r},t)\int d{\bf r'} \phi({\bf r'},t){\bf \nabla}U({\bf r-\bf{r'}})\right)$ which scales as $s^{\chi_1 +\chi_2 +1}$, determines the scaling of dynamical observables. This is because birth and apoptosis events (which occur on time scales $t>\frac{1}{k_b}$) are coupled to density evolution equation for $\phi({\bf r}, t)$ (see equation \ref{phi10} ). The birth-death term $k_a \phi^2$ in the cell density equation scales as $s^{2\chi_2}$, which determines the scaling for the CCs in the long-time limit.

\floatsetup[figure]{style=plain,subcapbesideposition=top}

%\begin{figure}[t]
%  \contcaption{Snapshot of tumor simulation with embedded tracers. {\bf (a)} A 3D simulated spheroid consisting of approximately 4800 CCs and 100 TPs. The CCs are in cyan, and the tracers are in red.  {\bf (b)} The spheroid shown above was rendered by making the CC cells transparent (light colored cyan) in order to show the interior of the spheroid. The TPs are opaque. Some of the TPs appear black because it is a depiction of a 3D image. The purpose of displaying these snapshots is to visually show that the TPs are randomly distributed within the multicellular spheroid, implying that their migration is largely determined by the forces arising from the CCs.}% Continued caption
%\end{figure}
%{\bf SS AND HS: The label Eq. S15 cuts into the equation. Please Fix. Thanks.}
\subsection{Stochastic Quantization}
{\color{blue}
%\textbf{Method of Calculation:}
{\color{black}
%In this section we outline the main features of the stochastic quantization method of
%Parisi and Wu~\cite{Parisi81ES,Damgaard87PhysRep,Himadri06PLA,Himadri06PRE}. 
A classical Langevin equation may be written as a path integral with an appropriate action. Similarly, the  path integral given by the distribution,
\begin{equation}
\exp[-\mathcal{S}(\Phi(x,t)]/\int D\Phi \exp[-\mathcal{S}(\Phi(x,t)],
\end{equation}
 can be described as the equilibrium limit of a statistical system that is coupled to a thermal reservoir~\cite{Parisi81ES,Damgaard87PhysRep,Himadri06PLA,Himadri06PRE}. The coupling to the thermal bath is simulated by letting $\Phi(x,t)$ ($t$ is real time) evolve in a fictitious time $\tau$ in the presence of stochastic noise subject to a systematic force $\propto \frac{\partial \Phi(x,t,\tau)}{\partial \tau}$. As $\tau \rightarrow \infty$, the distribution in the above equation would be recovered provided the stochastic noise satisfies the Fluctuation Dissipation Theorem (FDT). Here, $\mathcal{S}$ is a functional of the collective time-dependent density fields associated with the TPs and CCs.

Consider the evolution of  $\Phi(x,t,\tau)$ given by the Langevin equation,
\begin{equation}\label{2}
\frac{\partial \Phi(x,t,\tau)}{\partial \tau}=
-\frac{\delta \mathcal{S}}{\delta \Phi}+g(x,t,\tau)
\end{equation}
where $g$ is a Gaussian random variable with zero mean and,
\begin{equation}\label{3}
<g(x,t,\tau)g(x^{\prime},t^{\prime},\tau^{\prime})>=
2\delta(x-x^{\prime})\delta(t-t^{\prime})\delta(\tau-\tau^{\prime}).
\end{equation}
Properties of interest involving the density field,
$\Phi(x,t,\tau)$, can be evaluated by averaging over the Gaussian noise $g$, and  taking the limit of $\tau \rightarrow \infty$. For example,$<\Phi(x_{1},t_1)\Phi(x_{2},t_2)>$ can obtained using, 
\begin{eqnarray}\label{3}
\lim_{\tau \to\infty}<\Phi(x_{1},t_1,\tau)\Phi(x_{2},t_2,\tau)>_g &=&<\Phi(x_{1},t_1)\Phi(x_{2},t_2)>\\ \nonumber
&=&\frac{\int \mathcal{D}[\Phi]\Phi(x_{1},t_1)\Phi(x_{2},t_2)) 
e^{-S[\Phi]}}{\int \mathcal{D}[\Phi]e^{-S[\Phi]}}
\end{eqnarray}\\

We outline the steps to obtain the dynamic exponent $z$.\\
1. We consider a scalar field $\Phi({ k})$
in the momentum space satisfying the general equation of motion,
\begin{equation}\label{eqn10a}
\dot{\Phi}({k}) = -A({ k})\Phi ({k})-B(\Phi) + \eta_{\Phi}({ k}),
\end{equation}
where $B(\Phi)$ is a non-linear function of $\Phi$ and $\eta_{\Phi}(k) $ is a noise term 
with correlation $<{\eta_{\Phi}}({k}){\eta_{\Phi}}({ k}^{\prime})>
=2D_{0}\delta({ k}+{k}^{\prime})$. The probability distribution corresponding 
to the noise is given by,
%\begin{equation}\label{eq11a}
$P(\eta_{\Phi}) \propto \exp[ - \int \frac{d^{D}k}{(2\pi)^{D}}
\frac{dw}{2\pi}\frac{1}{4D_{0}}
\eta_{\Phi}(k,w) \eta_{\Phi}(-k,-w)]$.
%\end{equation}
In the momentum-frequency space, Eq.(\ref{eqn10a}) reads
\begin{equation}\label{eq12a}
[-iw + A(k)]\Phi(k,w)+B_{k,w}(\Phi) = \eta_{\Phi}(k,w)
\end{equation}
2. The probability distribution written in terms of $\Phi(k,w)$
instead of $\eta_{\Phi}(k,w)$ is,
\begin{eqnarray}\label{eq13a}
 P &\propto & \exp\{-\frac{1}{4D_{0}}\int_{k,w} 
\{[-iw+A(k)]\Phi(k,w)+B_{k,w}(\Phi)\}
\{[iw+A(-k)]\Phi(-k,-w)+B_{-k,-w}(\Phi)\}\}\nonumber\\ 
&=&\int \mathcal{D}[\Phi]e^{-S(\Phi)}\nonumber\\& &
\end{eqnarray}
3. Introduce a fictitious time $'\tau'$ and consider 
all the variables to be functions of $\tau$ in addition to ${k}$
and $w$. A Langevin equation in the $'\tau'$ variable is,
\begin{equation}\label{eq14a}
\frac{\partial \Phi(k,w,\tau)}{\partial \tau}=
- \frac{\delta S}{\delta \Phi(-k,-w,\tau)}
+g(k,w,\tau)
\end{equation}
with $<gg>=2\delta({k}+{k}^{\prime})\delta(w+w^{\prime})
\delta(\tau -\tau^{\prime})$.

4. In order to obtain the scaling laws for the mean square displacement (MSD) 
it suffices to work at arbitrary $\tau$. From Eq.(\ref{eq14a}), 
%that in the absence of the nonlinear terms (the terms involving $B(\phi)$), 
the Greens function $G$, is  
\begin{equation}\label{eq17a}
G^{-1}=-i\omega_{\tau} +\frac{w^{2}+A^{2}}{2D_{0}} +\Sigma (k,w,\omega_{\tau}),
\end{equation}
where $\omega_{\tau}$ is the frequency corresponding to the fictitious
time $\tau$. The nonlinear terms in $B$ contribute to the self-energy $\Sigma (k,w,\omega_{\tau})$ .
The correlation function is given by $C=\frac{1}{\omega_{\tau}}Im G$ through the FDT.

5. The nonlinear terms are treated using perturbation theory. 
A self consistent calculation yields the scaling exponents.  
The dimensionality of $\Sigma$ can be counted by using the dimensionality of $C$ through FDT relation $C=1/\omega_{\tau}ImG$.
If the 
frequency scale is  modified to $k^{z}$, 
then $G$ scales as $k^{-2z}$. We obtain the dynamic exponent $z$ by matching the power count of last two terms in Eq. (\ref{eq17a}). 

}

}

\subsection{Theory for  TP and CC dynamics}
%To understand the collective dynamics of the TPs, we use the Parisi-Wu~\cite{Parisi81ES} stochastic quantization method developed by in the context of quantum field theory. 
%{\color{blue}We exploit the principle of stochastic quantization to set up an alternative approach to the study of the coupled non-equilibrium dynamics of TP-CC mixtures. The advantage is that by introducing a fictitious time, we are able to bring back to fluctuation dissipation theorem and this allows a power counting scaling analysis for the binary system (TP-CC). Study of scaling solutions for both TPs and CCs is thereby greatly simplified.

%}

The evolution of  the CC density, described by Eq.~(\ref{phi10}) is an out of equilibrium process characterized by the absence of FDT. The usual analytic route to solve the problem is to introduce a response field $\tilde{\phi}$. Here, we need to calculate both the response function ($G=<\phi \tilde{\phi}>$) and correlation function ($C=<\phi \phi>$) separately because of the FDT is violated.
%The key advantage of the stochastic quantization method is that we do not need to calculate both the correlation and response functions. The FDT is satisfied in the fictitious time. The FDT relation enables us to obtain the scaling of the correlation function, once the scaling of the response function is known. By taking the infinite limit in fictitious time, one can obtain the correlation functions in real-time. The scaling solution of the problem can be obtained by power counting analysis instead of doing renormalization group calculation.

The density and noise fields are functions of $\tau$ in addition to the real parameters, $\bf{r}$ and $t$. 
The equation of motion for $\Psi({\bf r},t)$ in the $`\tau$' variable is,
\begin{equation}\label{langefic}
\frac{\partial \Psi({\bf k},w,\tau)}{\partial \tau}=-\frac{\delta \mathcal{S}}{\delta \Psi(-{\bf k},-w,\tau)}+g({\bf k},w,\tau) \, ,
\end{equation}
with $<g g>=2 \delta(k+k')\delta(w+w')\delta(\tau-\tau')$. {The field $\Psi$ represents both  $\psi_1$ and $\phi_1$ fields.} We are assured that the distribution for the density fields in Eq.  \ref{langefic} will approach $\exp[-\mathcal{S}(\phi_1, \psi_1)]$ in the $\tau\rightarrow \infty$ limit, because FDT is preserved in the $\tau$ variable. The action $\mathcal{S}(\phi_1, \psi_1)$ may be obtained by writing the joint probability distribution $P(\eta'_{\phi_1},\eta'_{\psi_1}) \propto \text{exp}[-\int \frac{d^d{\bf k}}{(2\pi)^d}\frac{dw}{2\pi}\{\frac{1}{2(k_a \phi_0+k_b \phi_0^2+D_{\phi} \phi_0 k^2)}\eta'_{\phi_1}({\bf k},w)\eta'_{\phi_1}(-{\bf k},-w)+\frac{1}{(2D_{\psi} \psi_0 k^2)}\eta'_{\psi_1}({\bf k},w)\eta'_{\psi_1}(-{\bf k},-w)\} ]$ corresponding to the noise terms $\eta_{\phi_1}$ and $\eta_{\psi_1}$ associated with the CC (Eq.(\ref{phi10})) and TP equations (Eq.(\ref{trdensity})), respectively. 
The noise correlations are given by $<\eta'_{\phi_1}({\bf k},w)\eta'_{\phi_1}(-{\bf k},-w)>=2(k_a \phi_0+k_b \phi_0^2+D_{\phi} \phi_0 k^2)$, and $<\eta'_{\psi_1}({\bf k},w)\eta'_{\psi_1}(-{\bf k},-w)>=2D_{\psi} \psi_0 k^2$.
The action $\mathcal{S}(\phi_1, \psi_1)$ in terms of $\phi_1({\bf k},w)$ and $\psi_1({\bf k},w)$ may be calculated using Eq.(\ref{phi10}) and Eq.(\ref{trdensity}). The expression is too complicated to reproduce here, and is not needed for obtaining the main results. For the simpler case (for a single $\phi$ field) we have derived it elsewhere~\cite{Himadri18PRE}.

%{\bf THE FOLLOWING HAS BEEN STATED PREVIOUSLY. SO I PROPOSE REMOVING IT.}
We obtain the following self-consistent equation for the
self-energy $\Sigma_{\psi_1}({\bf k},\omega, \omega_{\tau})$ from the calculation of response function: 
\begin{equation}\label{scale1}
\Delta \nu =\frac{D}{2\nu }\Sigma_{\psi_1}({\bf k},\omega, \omega_{\tau})
\end{equation}
where, $\nu=D_{\psi} k^2 +\psi_0 k^2 U({\bf k})
+\phi_0 k^2 U({\bf k})$ and $D=2D_{\psi} \psi_0 k^2$. {\color{black}Physically the self-energy term determines the contribution to the relaxation rate arising from the non-linearity.} {\color{black} We can carry out
the momentum count of Eq.(\ref{scale1}), keeping in mind that 
$\Delta\nu \sim k^{z}$, to extract the dynamic exponent $z$ in different time regimes. 
%We follow the procedure for obtaining the scaling laws for the systems far from equilibrium, illustrated in the previous section. {The dynamics in Eq.(\ref{langefic}) requires only the calculation of the response function ({$G$}) as the correlation function ({$C$}) is related to response function through FDT relation, which in Fourier space may be written as},
%\begin{equation}
%C=\frac{1}{\omega_{\tau}}\text{Im} G.
%\end{equation}

%In order to derive the scaling laws for the MSD it suffices to work at arbitrary but not necessarily $\tau \rightarrow \infty$ limit. 

%{In order to test the predictions, we also simulated a three dimensional tumor spheroid with embedded TPs. }

\section{Simulation Models:} {In order to test the theoretical predictions and determine the mechanisms underlying the unusual dynamics, we simulated a three dimensional tumor spheroid with embedded TPs. }{An agent based model~\citep{Abdul17Nature, malmi2019dual, sinha2019spatially} is used for the tumor spheroid. The cells  are treated as deformable objects. The size of the CCs increase  with time as the tumor grows, and divide into two identical cells upon reaching a critical mitotic radius ($R_m$). The mean cell cycle time is $\tau$. The cell cycle time $\tau$ is expressed in units of $\tau=15~hrs$. The CCs can also undergo apoptosis. As in the theory, the TPs are  inert, and their sizes and the number are constant throughout the simulations.} 
We include CC-CC, CC-TP and TP-TP interactions. We use two potentials for the interactions.% in order to assess the effect of the form of the potential on the dynamics of the TPs and CCs.

\subsection{Hertz potential} The form of the Hertz forces between the CCs is the same as in previous studies \cite{Abdul17Nature, malmi2019dual,drasdo2005single, schaller2005multicellular, pathmanathan2009computational}. The physical properties of the CC, such as the radius, elastic modulus, membrane receptor and ligand concentration characterize the strength of the inter-cellular interactions. The elastic forces between two spheres with radii $R_i$ and $R_j$, is given by,
\begin{equation}
F_{ij}^{el}=\frac{h_{ij}^{\frac{3}{2}}}{\frac{3}{4}(\frac{1-\nu_i^2}{E_i}+\frac{1-\nu_j^2}{E_j})(\sqrt{\frac{1}{R_i}+\frac{1}{R_j}})},
\label{hertzrepul}
\end{equation}
where $E_i$ and $\nu_i$ are, respectively, the elastic modulus and Poisson ratio of the $i^{th}$ cell. Since, the CCs or the TPs are deformable, the elastic force depends on the overlap, $h_{ij}$, between two cells.
The adhesive force, $F_{ij}^{ad}$, between the CCs is proportional to the area of contact ($A_{ij}$) \cite{palsson2000model}, and is calculated using, \cite{schaller2005multicellular},
\begin{equation}
F_{ij}^{ad}=A_{ij}f^{ad}\frac{1}{2}(c^{rec}_i c^{lig}_j + c^{rec}_j c^{lig}_i ),
\label{hertzattra}
\end{equation}
where $c_i^{rec} (c_i^{lig})$ is the receptor (ligand) concentration on the surface of the cells, and are taken to be unity in the present study. The coupling constant $f^{ad}$ allows us to scale the adhesive force to account for variable receptor and ligand concentrations.

Repulsive and adhesive forces in Eqs.(\ref{hertzrepul}) and (\ref{hertzattra}) act along the unit vector $\vec{n}_{ij}$ pointing from the centers of cells $j$ and $i$. Therefore, {the net force on} cell $i$ ($\vec{F}_{i}^{H}$) is given by the sum over its nearest neighbors [NN(i)],
\begin{equation}
\vec{F}_{i}^{H} = \Sigma_{j \epsilon NN(i)}(F_{ij}^{el}-F_{ij}^{ad})\vec{n}_{ij}.
\label{hertzian_force}
\end{equation}

To model the TP-TP and TP-CC interactions, we assume that the TPs are CC-like objects, which mimics experiments \cite{Dolega17NC}. Therefore, CC-TP and TP-TP interactions are the same as CC-CC interactions.
%As in the theory both $F_{ij}^{el}$ and $F_{ij}^{ad}$ are short range forces because they depend only on $h_{ij}$, the overlap between CCs. 

\subsection{Gaussian potential} In the theoretical treatment, we assumed that the CC-CC interaction is given by a sum of Gaussian terms (Eq.~(\ref{potential})). 
%\begin{multline}
%U({\bf r}(i)-{\bf r}(j))=\frac{\nu}{(2\pi\lambda^2)^{3/2}}e^{\frac{-|{\bf r}(i)-{\bf r}(j)|^2}{2\lambda^2}}  \\
% -\frac{\kappa}{(2\pi\sigma^2)^{3/2}}e^{\frac{-|{\bf r}(i)-{\bf r}(j)|^2}{2\sigma^2}}
%\end{multline}
%where, $\lambda$ and $\sigma$ are the range of repulsive and attractive interactions, and $\nu$ and $\kappa$ denote the strength of the repulsive and attractive interactions. 
For this potential, the force ${\bf F}_{ij}^{G}$ on cell $i$, exerted by cell $j$, is,
\begin{equation}
{\bf F}_{ij}^{G}=\frac{1}{(2\pi)^{3/2}}[\frac{\nu e^{\frac{-r^2}{2\lambda^2}}}{\lambda^5}-\frac{\kappa e^{\frac{-r^2}{2\sigma^2}}}{\sigma^5}]{\bf r}
\label{gaussian_force}
\end{equation}
where ${\bf r}$ is ${\bf r}(i)-{\bf r}(j)$. 
We write $\lambda$ and $\sigma$ as $\lambda=\widetilde{\lambda}(R_i+R_j)$ and $\sigma=\widetilde{\sigma}(R_i+R_j)$, as the ranges of interactions corresponding to the repulsive and attractive interactions, respectively. {In our simulations, the CCs grow and divide, their radii change in time, and therefore $\lambda$ and $\sigma$ also change in time. However, since these interactions are short-ranged, we assume them to be constant (as done in the theory). So, we fixed $\lambda=\widetilde{\lambda}(2R_d)$ and $\sigma=\widetilde{\sigma}(2R_d)$, where $R_d$ ($\approx 4 \mu m$) is the size of a daughter cell (introduced in the next section)}. For simplicity, we write force ${\bf F}_{ij}^{G}=[\frac{\widetilde{\nu} e^{\frac{-r^2}{2\lambda^2}}}{\lambda^2}-\frac{\widetilde{\kappa} e^{\frac{-r^2}{2\sigma^2}}}{\sigma^2}]{\bf r}$, where $\widetilde{\nu}=\frac{1}{(2\pi)^{3/2}}\frac{\nu}{\lambda^3}$ and $\widetilde{\kappa}=\frac{1}{(2\pi)^{3/2}}\frac{\kappa}{\sigma^3}$. 
 In the simulations, we fixed $\widetilde{\nu}=0.03$, {$\widetilde{\lambda}=0.28$}, $\widetilde{\kappa}=0.003$ and $\widetilde{\sigma}=0.4$.
 
\subsection{Equation of Motion}  The equation of motion governing the dynamics of TP and CCs is taken to be, 
\begin{equation}
\label{eqforce}
\dot{\vec{r}}_{i} = \frac{\vec{F}_{i}}{\gamma_i},
\end{equation}
where $\dot{\vec{r}}_{i} $ is the velocity of $i^{th}$ CC or TP, $\vec{F}_{i}$ is the force on $i^{th}$ CC/TP (see equation \ref{hertzian_force} and \ref{gaussian_force}), and $\gamma_i$ is the damping term { (for details see reference \cite{Abdul17Nature})}. 

\subsection{Cell division and Apoptosis} The CCs are either dormant or in the growth phase depending on the value of the pressure. The pressure on cell $i$ ($p_i$) due to $NN(i)$ neighboring cells is calculated using the Irving-Kirkwood equation, 
\begin{equation}
\label{pressure}
p_{i} =  \frac{1}{3V_i}\Sigma_{j \epsilon NN(i)}  {\bf F}_{ij}\cdot d{\bf r}_{ij},
\end{equation}
where ${\bf F}_{ij}$ is the force on the i$^{th}$ cell due to j$^{th}$ cell and $d{\bf r}_{ij} = {\bf r}_i- {\bf r}_j$. The volume of the $i^{th}$ cell ($V_i$) is $\frac{4}{3}\pi R_i^3$, where $R_i$ is the radius of the i$^{th}$ cell.  If $p_{i}$ exceeds a pre-assigned critical limit $p_c$ ($= 1.7\times 10^{-6} $ MPa) the CC enters a dormant phase. The dormancy criterion serves as a source of mechanical feedback, which limits the growth of the tumor spheroid \cite{shraiman2005mechanical,alessandri2013cellular,conger1983growth,puliafito2012collective,gniewek2019biomechanical}. The volume of a growing cell increases at a constant rate, $r_V$. The cell radius is updated from a Gaussian distribution with the mean rate $\dot{R} = (4\pi R^2)^{-1} r_V$. Over the cell cycle time $\tau$, 
\begin{equation}
r_V = \frac{2\pi (R_{m})^3}{3\tau},
\end{equation}
where $R_{m}$ is the mitotic radius. A cell divides once it grows to the fixed mitotic radius. To ensure volume conservation, upon cell division, we use $R_d = R_{m}2^{-1/3}$ as the radius of the daughter cells. The resulting daughter cells are placed at a center-to-center distance $d = 2R_{m}(1-2^{-1/3})$ (Fig. 1 in the main text). The direction of the new cell location is chosen randomly from a uniform distribution on a unit sphere.

We initiated the simulations with 100 TPs and 100 CCs. The coordinates of the CCs and TPs were sampled using a normal distribution with mean zero and standard deviation $50~\mu m$. The initial radii of the CCs and TPs were sampled from a normal distribution with mean $4.5~\mu m$, and a dispersion of  $0.5~\mu m$. %Hence, the CCs and TPs are polydisperse cells. 

%\begin{figure}[h]
%\vspace{-4.3 in}
	%\includegraphics[width=.90\textwidth]{selfenergy_tracer.pdf}
	%\vspace{-1.1 in}
	%\caption{Diagrammatic representations of the self-energy terms contributing to sub-diffusive (a), super-diffusive (b) and hyper-diffusive (c) behaviors. The solid line represents the full Greens function for the tracer and the dotted line represents the full Greens function for the cell. The line with a crossed circle represents the full correlation function. }  
%	\label{fig:rg61}
%\end{figure}

\section{ Details of the Results}
\subsection{{TPs exhibit sub-diffusive dynamics in the intermediate time regime }}
 { At $t<\frac{1}{k_b}$, the non-linear terms ${\bf \nabla }\cdot \left(\psi_1({\bf r},t)\int d{\bf r'} \psi_1({\bf r'},t){\bf \nabla}U({\bf r-\bf{r'}})\right)$ and
  ${\bf \nabla }\cdot \left(\psi_1({\bf r},t)\int d{\bf r'} \phi_1({\bf r'},t){\bf \nabla}U({\bf r-\bf{r'}})\right)$ describing the TP-TP and TP-CC interactions respectively,  govern the scaling behavior of $\Delta_{TP}(t)$, the MSD.}~In the spirit of self-consistent mode coupling theory, we replace $\nu$ by $\Delta \nu$ in the self-energy term $\Sigma(k, \omega , \omega_{\tau} )$ (Eq.(\ref{scale1})). 
{Using the scale transformation, we find $\omega \sim k^z$, $\omega_\tau \sim k^{2z-2}$, $G_{\psi_1} \sim k^{-2z+2}$, $C_{\psi_1} \sim k^{-4z+4}$, and the vertex factor $V \sim k^{z}$. The relevant part $V$ is $\frac{1}{(D_{\psi_1} \psi_0 k^2)}(\{i \omega+D_{\psi_1}k^2+\psi_0 k^2 U({\bf k})\} \{(-{\bf k'} \cdot {\bf k}) U({\bf k'})\})$.
%+ \{i \omega'+D_{\psi_1}k'^2+\psi_0 k'^2 U({\bf k'})\} \{(-{\bf k'}\cdot {\bf k}) U(-{\bf k})\} +\{i \omega'+D_{\psi_1}k'^2+\psi_0 k'^2 U({\bf k'})\}  \{(-{\bf k'} \cdot ({\bf k}-{\bf k'})) U({\bf k}-{\bf k'})\})$. 
The self energy term 
%(shown in fig.(\ref{fig:rg61}a)) 
has the structure:
$\Sigma({\bf k},\omega, \omega_{\tau})\sim  \int \frac{d^d {\bf k'}}{(2\pi)^d} \frac{d\omega'}{2\pi} \frac{d\omega'_\tau}{2\pi} V V GC$.}
By carrying out the momentum count in $\Sigma({\bf k},\omega, \omega_{\tau})$, and noting  that $\Delta \nu \sim k^z$, we find $\Sigma({\bf k},\omega, \omega_{\tau})\sim k^{d-z+4}$. 
Using Eq. (\ref{scale1}), we obtain $k^{z}\sim k^{d-z+4}$, which leads to $z=2+\frac{d}{2}$.

The MSD of the TPs scales with $t$ as, 
\begin{equation}
<[r(t)-r(0)]^2>\sim t^{2/z}=t^{\alpha_{TP}}.
\end{equation}
In the intermediate times, $\beta{TP}=\frac{2}{z}= 0.57$, a value that holds provided the interaction potentials involving the CCs and TPs are given by Eq. \ref{potential}.
%and $\alpha=\frac{4}{8}= 0.5$ for interaction potential $U_1=U_0/\cosh^2(r/a)$, implying sub-diffusive behavior. %Although sub-diffusive behavior is preserved at intermediate times, 
%The scaling exponents depend on the form of interaction potential, which shows that the behavior of MSD is non-universal in the intermediate time limit. 
%{\bf GUYS: GIVEN THAT THIS PAPER IS LONG AND HEAVY DO WE NEED THE FOLLOWING DEALING WITH A DIFFERENT POTENTIAL?}
%On the other hand, if the TP-TP interaction is $U_1=U_0/\cosh^2(r/a)$ instead, we obtain $\beta^T_{TP}=\frac{4}{8}= 0.5$. 
%The reason for the dependence on the interactions is that the non-linear term $\int d{\bf q} (-{\bf q}\cdot {\bf k})U({\bf q})\psi_1({\bf q})\psi_1({\bf k}-{\bf q})$ involving the TP-TP interactions, determines the scaling exponent in the intermediate time scale. For the Gaussian potential, $U({\bf q})=\exp[-q^2/2\sigma^2]$, which has no explicit $q$ dependence. For the potential of the form  
%$U_1=U_0/\cosh^2(r/a)$, $U({\bf q})=a^2 q \sqrt{\pi/2} \cosh(a q \pi/2)$, i.e., $U({\bf q})$ has explicit $q$ dependance, which is reflected in the different scaling exponents for different potential form. In this case $\beta^T_{TP}=4/8=0.5$. Regardless of the nature of the potentials, as long as they are short-ranged, the intermediate value of $\beta_{TP}$ is less than unity, indicating glass-like dynamics in the intermediate time scales ($t~\textless~k_b^{-1}$).

\subsection{ Hyper-diffusion of the TPs in the long time limit}} At long times ($t \gg k_b^{-1}$), the effects of non-linearity in the TP-CC interactions together with non-equilibrium forces due cell division and apoptosis, determine the TP dynamics. 
%The corresponding self-energy term in Eq.(\ref{scale1}) is shown in Fig.(\ref{fig:rg61}c). 

%{\bf PLEASE DEFINE $V_1$, $V_2$, and $V_3$ more clearly}
As in the previous section, we use the scale transformations: $\omega \sim k^z$, $\omega_\tau \sim k^{2z-2}$, $G_{\psi_1} \sim k^{-2z+2}$, $C_{\psi_1} \sim k^{-4z+4}$, $G_{\phi_1} \sim k^{-2z}$, $C_{\phi_1} \sim k^{-4z}$, and vertex factors $V_1 \sim k^{2}$, and $V_{2} \sim V_3 \sim k^{z}$. The relevant part of $V_1$ is  $\frac{1}{2(k_a \phi_0 +k_b \phi_0^2)} k_b ({\bf q} \cdot {\bf k}) U({\bf q})$, and the form of $V_2$ or $V_3$ is $\frac{1}{(D_{\Psi} \Psi_0 k^2)}\{i \omega+D_{\Psi}k^2+\Psi_0 k^2 U({\bf k})\} \{(-{\bf k'} \cdot {\bf k}) U({\bf k'})\}$. Noting that  $\Delta \mu \sim k^z$, we find $\Sigma({\bf k},\omega, \omega_{\tau})\sim  \int \frac{d^d {\bf k'}}{(2\pi)^d} \frac{d\omega'}{2\pi} \frac{d\omega'_\tau}{2\pi} V_1 V_2 V_3 G_{\psi_1} G_{\psi_1} C_{\phi_1} C_{\psi_1} \sim k^{d+4-7z}$. From the self-consistent equation (Eq. (\ref{scale1})) we find that the dynamic exponent, $z=(d+4)/8$. The value of  $\alpha{TP} =\frac{2}{z}=\frac{16}{7}=2.28$. Thus, we predict that the TPS must exhibit {\it hyper-diffusive} motion at long times. 
%The self energy term $\Sigma({\bf k},\omega, \omega_{\tau_f})$ (Fig(\ref{fig:rg61}c)) depends linearly on the birth rate for the CCs. The different values for the cell cycle time, i.e., the different values of birth rate ($k_b$), do not change the scaling of the self energy term. It only changes the coefficients of the self energy term. Therefore, the MSD exponent for the TPs is independent of the cell cycle time as long as $t$ is large.

\floatsetup[figure]{style=plain,subcapbesideposition=top}
\begin{figure}
\subfloat[]{\includegraphics[width=0.6\linewidth] {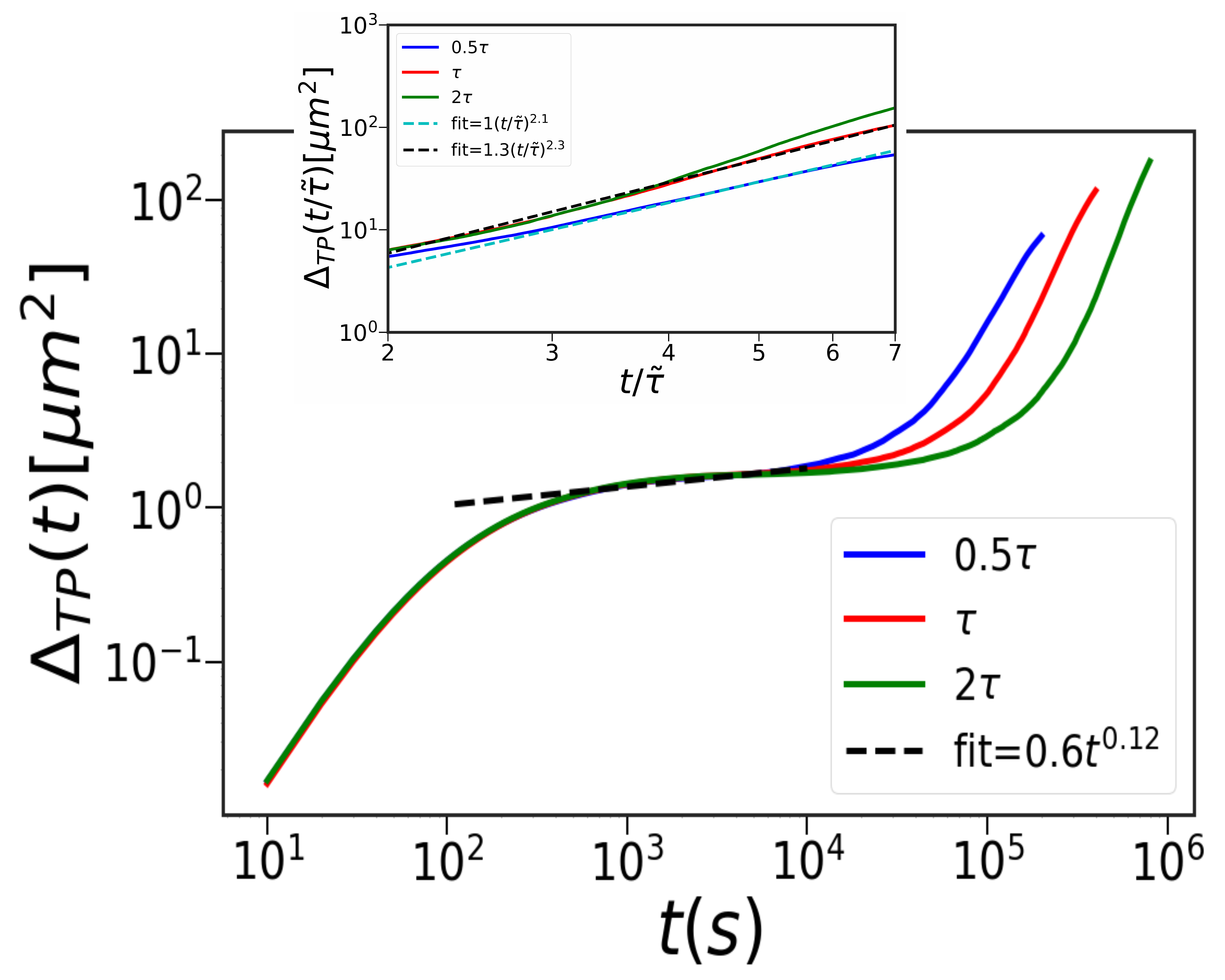}\label{fig1a}}
\par
\subfloat[]{\includegraphics[width=0.6\linewidth] {tracer_msd_gaussian_diff_birth_ik_new.pdf}\label{fig1b}} 

\caption{  {\color{black} MSD, ($\Delta_{TP}(t)$), of the TPs as a function of time ($t$) for the two types of CC interactions. {\bf(a)} $\Delta_{TP}$  calculated using Eq. $\ref{hertzian_force}$ for the forces between the CCs and TPs. The curves are for 3 cell cycle times (blue -- $0.5\tau$, red --  $\tau$, and green -- $2\tau$). Time taken to reach the super-diffusive regime, which is preceded by a sub-diffusive regime, increases as cell cycle time increases. In the long-time, $\Delta_{TP}(t)$ undergoes hyper-diffusive motion ($\Delta_{TP}\sim t^{\alpha_{TP}}$  with $\alpha_{TP}>2$), which is highlighted in the inset. The x-axis of the inset plot is scaled by $\frac{1}{\tilde{\tau}}$. The black (cyan) dashed line shows exponent $\alpha_{TP}=2.30$ (2.1). The curve with 0.5 $\tau$ is best fit best with $\alpha_{TP} =2.1$. {\bf(b)} Same as (a) except the Gaussian potential (equation \ref{gaussian_force}) is used in the simulations. Interestingly, $\alpha_{TP}$ does not change appreciatively.}}
\label{tracer_msd_birth}
\end{figure}

{\subsection{k-dependent diffusion coefficients}
One of our key findings is that the motion of the TPs is hyper-diffusive whereas the CC dynamics is super diffusive ($\alpha_{TP} > \alpha_{CC}$), whether the TPs are present or not.  For a homogeneous system, the density fluctuation obeys the Eq.(\ref{trdensity}) without the non-linear terms. Thus, the equilibrium fluctuations, ($< \psi_1(k,t) \psi_1(k,0)>=\psi_0~ \exp[-D k^2 t]$) decay exponentially. The diffusion-coefficient ($D$) is a constant, and the MSD exponent is unity. In this case, the relaxation time ($(D k^2)^{-1} \approx k^{-z}$) with $z=2$.  Deviation from this standard situation is suggestive of anomalous diffusion.  Systematic interactions and non-equilibrium forces between the CCs and TPs modify the density fluctuations, giving rise to an effective $k$-dependent diffusion coefficient.

\floatsetup[figure]{style=plain,subcapbesideposition=top}
\begin{figure}
{\includegraphics[width=0.7\linewidth] {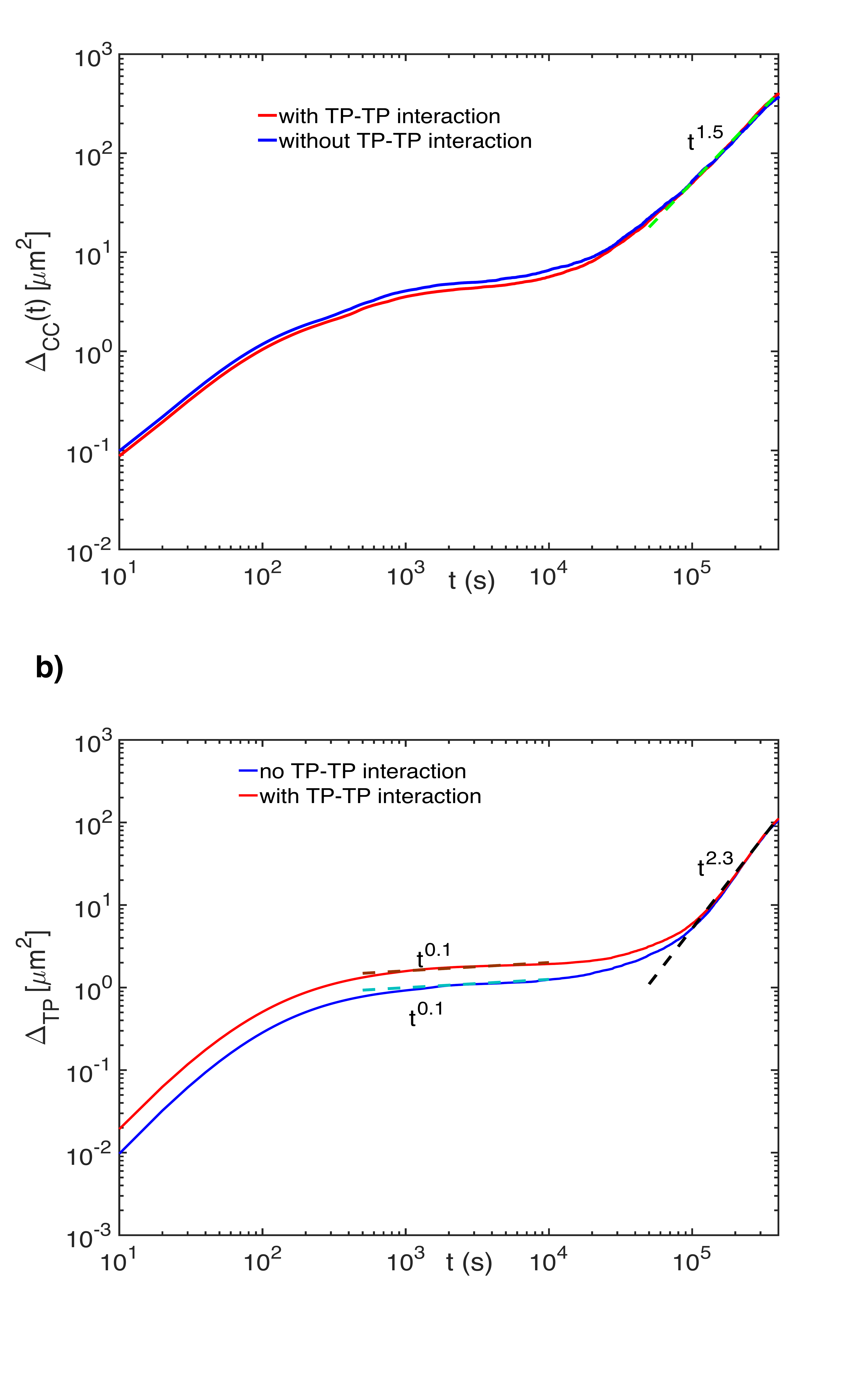}}
	%\par
\vspace{-.3in}
%\sidesubfloat[]{\includegraphics[width=0.8\linewidth] {tracer_msd_gaussian_diff_radius.eps}\label{fig3b}} 
\caption{{\color{black} {\bf TP-TP interactions do not affect the long time dynamics of TPs or CCs.} (a) $\Delta_{CC}$ with (red curve) and without (blue) TP-TP interactions. The TP-TP interactions plays no role in the CC dynamics. The cyan dashed line shows $\alpha_{CC}=1.5$ for both the cases. (b)  $\Delta_{TP}$ with (red curve) and without (blue) TP-TP interaction. $\Delta_{TP}$ differs in magnitude in the intermediate time. However, TP-TP interactions plays no role in the long time dynamics of the TPs.  }} .
\label{with_without}
\end{figure}

 Depending on the value of $z$, the TPs and CCs could exhibit sub or super or hyper diffusive motion. The non-linear term in Eq. (\ref{trdensity}) for the TP-TP interactions renormalizes the diffusion coefficient $D$. 
The effective TP diffusion constant, $D_{TP} \sim k^{z-2} =k^{-9/8}$, and for CCs the diffusion coefficient $D_{CC} \sim k^{-5/8}$.  %therefore the degree of diffusion coefficient indicates the higher degree of MSD exponent in TPs compared to CCs. 
The relaxation time for the dynamic structure factor for TPs (~$k^{-7/8}$) is small compared to the relaxation time for CCs (~$k^{-11/8}$), leading to a higher degree of anomaly in the diffusion of the TPs.}

\subsection{{$\alpha_{TP}$ is nearly independent of  the TP size}} 
\begin{figure}
{\includegraphics[width=0.7\linewidth] {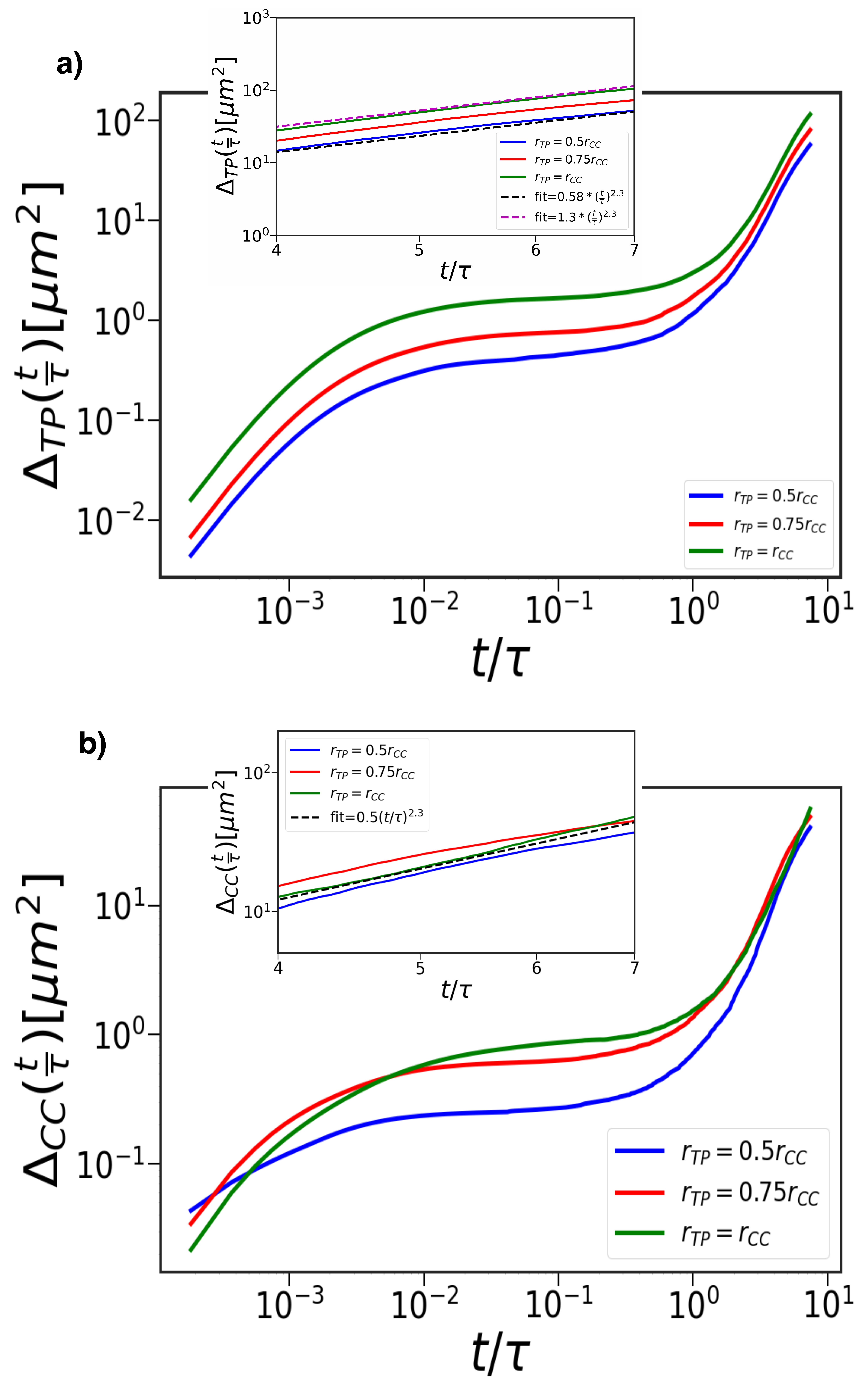}\label{fig3a}}
	%\par
\vspace{-.4in}
%\sidesubfloat[]{\includegraphics[width=0.8\linewidth] {tracer_msd_gaussian_diff_radius.eps}\label{fig3b}} 
\caption{{\color{black} Dependence of the TP size $\Delta_{TP}$ as a function of $t$. {\bf (a)} Data are for the Hertz potential (see equation \ref{hertzrepul}-\ref{hertzian_force}). {Time is scaled by $\tau$}. From top to bottom, the curves correspond to decreasing TP radius ($r_{TP} = r_{CC}$ (green), $r_{TP} = 0.75r_{CC}$ (red) and $r_{TP} = 0.5r_{CC}$ (blue), where $r_{CC} =4.5 \mu m$ is average CC radius). TPs with larger radius have larger MSD values  in the intermediate time ($\frac{t}{\tau}\leq \mathcal{O}(1)$). In the inset, we focus on the hyper-diffusive regime. The black and magenta dashed line serves as a guide to the eye with  $\alpha_{TP}=2.3$  {\bf (b)} Same as (a) but with the Gaussian interactions.}} .
\label{tracer_msd_diff_size}
\end{figure}

{We varied the radius of the TP ($r_{TP}$) from $0.5r_c$ to $r_c$, where $r_c=4.5 \mu m$ is the average CC radius.} Figure \ref{tracer_msd_diff_size} shows $\Delta_{TP}(t)$ as function of $t$ for the Hertz potential (Eq.~(\ref{hertzian_force})). Similar behavior is found for Gaussian potential as well. In the intermediate time regime, TPs with larger radius have higher MSD because they experience large repulsive forces due to higher excluded volume interactions. In the long time limit, $\Delta_{TP}$ exhibits hyper-diffusion (insets of Figure \ref{tracer_msd_diff_size}).{ The CC-TP interaction term,${\bf \nabla }\cdot \left(\psi_1({\bf r},t)\int d{\bf r'} \phi_1({\bf r'},t){\bf \nabla}U({\bf r-\bf{r'}})\right)$, in Eq.(\ref{trdensity}) shows that the radius only alters the interaction strength, and does not fundamentally alter the scaling behavior. The conclusion that the values of $\alpha_{TP}$ do not change, anticipated on theoretical grounds, is further supported by simulations.
 }

\floatsetup[figure]{style=plain,subcapbesideposition=top}

%\begin{figure}[t]
  %\contcaption{, where $r_c =4.5 \mu m$ is the average cell radius). TPs with bigger radius have larger MSD in the intermediate time ($\frac{t}{\tau}\leq \mathcal{O}(1)$). Two dashed lines, orange and red, serve as guide to eye the. The  exponent $\alpha_{CC}=2.26$. {\bf(b)} Same as a, except for the Gaussian potential \ref{gaussian_force}}% Continued caption
%\end{figure}

\floatsetup[figure]{style=plain,subcapbesideposition=top}
\begin{figure}
\subfloat[]{\includegraphics[width=0.62\linewidth] {cell_msd_diff_tracer_radius_ik_new.pdf}\label{fig5a}}
\par
\subfloat[]{\includegraphics[width=0.62\linewidth] {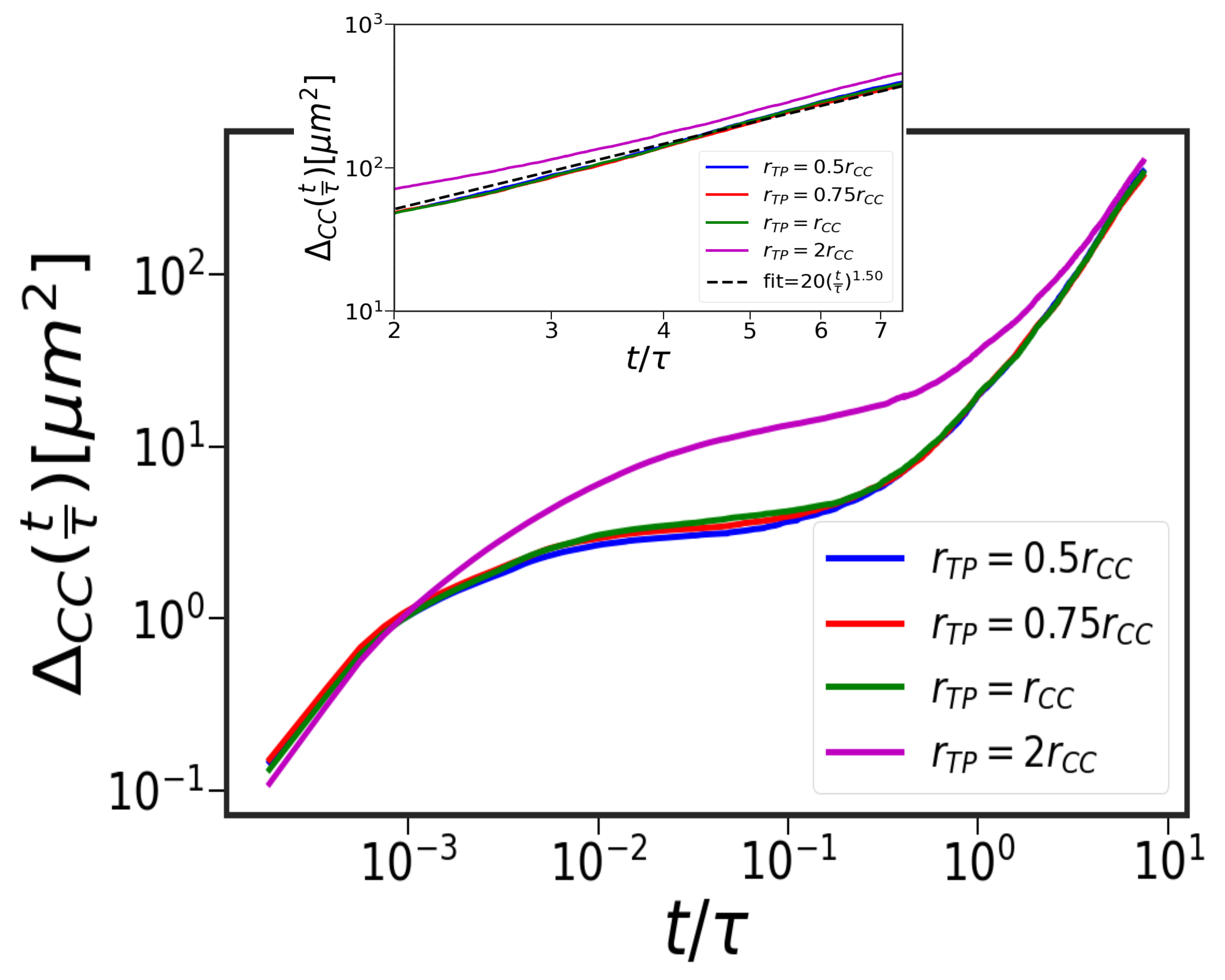}\label{fig5b}} 
\caption{{\color{black} Influence of TPs on CC dynamics using two potentials. $\Delta_{CC}$ as function of $t$ using the Hertz potential (equation \ref{hertzian_force}). From top to bottom, the curves are for different values of the radius of the TPs (magenta $r_{TP} = 2 r_{CC}$, green $r_{TP} = r_{CC}$, red $r_{TP} = 0.75r_{CC}$  and blue $r_{TP} = 0.5r_{CC}$ (appears to be hidden), where $r_{CC} =4.5~\mu m$ is the average cell radius). In the intermediate times, $\Delta_{CC}(t)$ is larger for TPs with larger radius. The inset focuses on the long time regime ($\frac{t}{\tau}>1$). The black line is a guide to show the value of  $\alpha_{CC}=$ 1.47. (b) Same as (a) but with the Gaussian potential.}}
\label{fig4aa}
\end{figure}

%\begin{figure}[t]
%  \contcaption{}% Continued caption
%\end{figure}

\subsection{{Influence of the TPs on CC dynamics}}
{\color{black} We calculated $\Delta_{CC}(t)$ using the Hertz (Gaussian) potential as a function of  the TP radius (Figure \ref{fig5a} (\ref{fig5b}). For both the potentials, the values of $\Delta_{CC}(t)$ for $t< \tau$ is greater as the TP sizes increase}.
%Figure \ref{fig4aa} shows the mean square displacement for the CC with varying TP radius. We observe that CCs that are influenced by the bigger TPs have higher MSD in the intermediate time regime. 
{Before cell division the number of CCs and TPs are similar, which explains the modest influence of the TPs on the dynamics of CCs in the intermediate time regime}. The larger TPs undergo stronger repulsion (the repulsive interaction is proportional to $R^2$) initially, which increases the magnitude of $\Delta_{CC}$. The long-time dynamics is not significantly affected by the CC-TP interactions (see Figure \ref{fig4aa}). In the absence of the TPs, the CCs exhibit super-diffusion where the MSD scales as $t^{\alpha_{CC}}$ with $\alpha_{CC} =1.33$. In the presence of the TPs, the CC dynamics remains super-diffusive with $\alpha_{CC} \approx 1.45$. %The simulation result agrees well with the theoretical prediction in the long-time limit. 

\subsection{{Straigtness Index}}
\begin{figure}
{\includegraphics[width=0.8\linewidth] {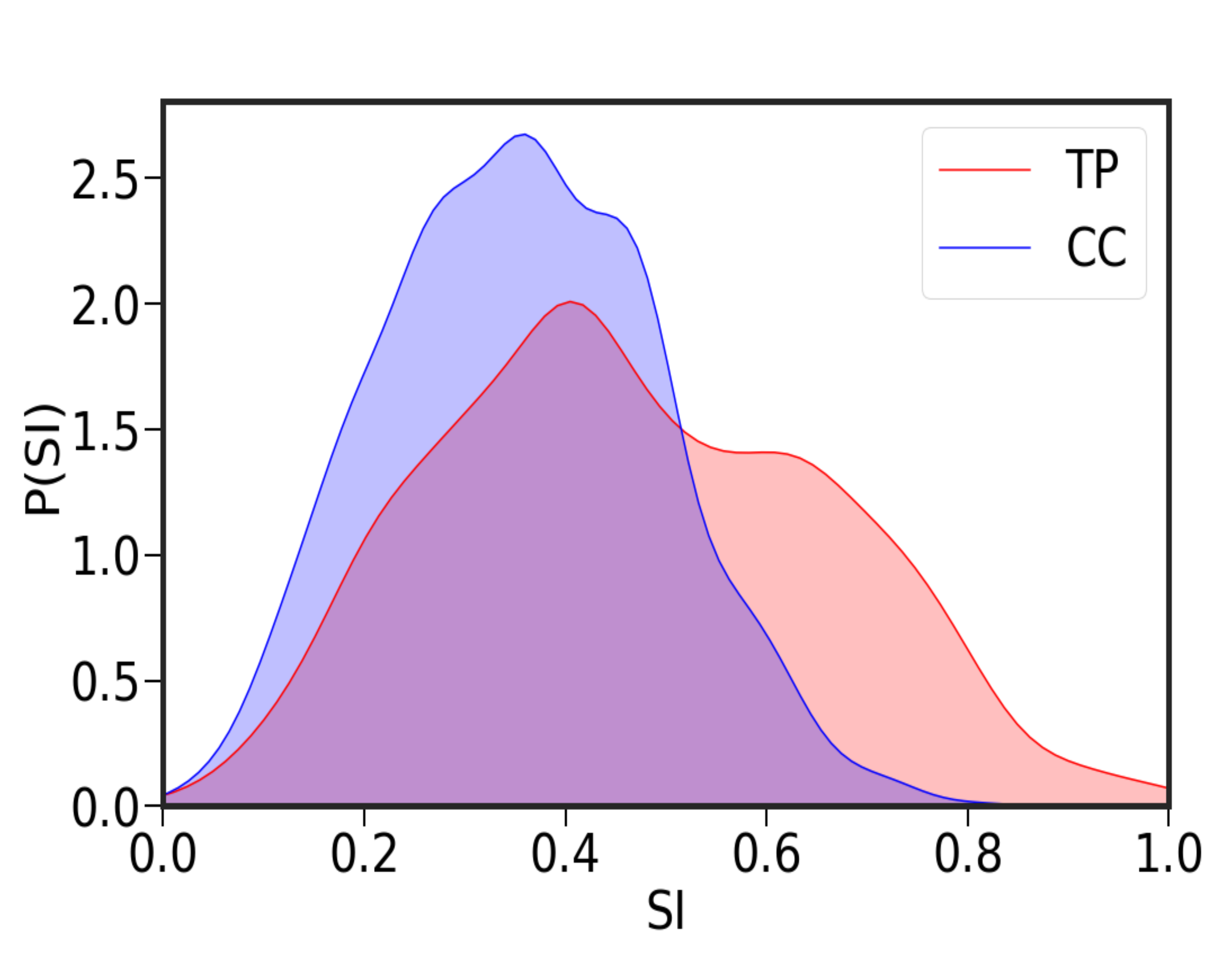}}
\caption{Distribution of the Straightness Index (SI). The red (blue) plot  for the TPs (CCs) shows that the TP trajectories are more rectilinear than the  CCs.}
\label{SI}
\end{figure}

The Straightness Index (SI) is, $SI_i=\frac{{\bf r}_i(t_f)-{\bf r}_i(0)}{\sum \delta {\bf r}_i(t)}$. The numerator ${\bf r}_i(t_f)-{\bf r}_i(0)$ is the net displacement of $i^{th}$ TP or CC, and the denominator $\sum \delta {\bf r}_i(t)$ is the total distance traversed. Fig. \ref{SI} shows that the TP trajectories are more straight or persistent over longer duration. During each cell division, the CCs are placed randomly causing the  trajectories to be less persistent, thus explaining the decreased persistence in their motion.

\floatsetup[figure]{style=plain,subcapbesideposition=top}

\clearpage
\bibliography{tracer.bib}
\bibliographystyle{unsrt}
%\bibliography{SI-cancer.bib}